\documentclass[useASM,usenatbib]{mn2e}
\usepackage{url}
\usepackage{color}
\usepackage[colorlinks=true,citecolor=blue]{hyperref}
\usepackage{graphicx,graphics,color}
\usepackage[pagewise]{lineno}
\usepackage[T1]{fontenc}
\usepackage{ae,aecompl}

\newcommand\chandra{{\it Chandra}}

\newcommand\kev{{\rm~keV}}

\newcommand\kms{\ifmmode {\rm~km\ s}^{-1} \else ~km s$^{-1}$\fi}
\newcommand\Hunit{\ifmmode {\rm~km\ s}^{-1}\ {\rm Mpc}^{-1}
        \else ~km s$^{-1}$ Mpc$^{-1}$\fi}
\newcommand\ctssec{\ifmmode {\rm~count\ s}^{-1} \else ~count s$^{-1}$\fi}
\newcommand\ergsec{\ifmmode {\rm~erg\ s}^{-1} \else
        ~erg s$^{-1}$\fi}
\newcommand\funit{\ifmmode {\rm~erg\ s}^{-1}\;{\rm cm}^{-2} \else
        ~ergs s$^{-1}$ cm$^{-2}$\fi}
\newcommand\phflux{\ifmmode {\rm~photon\ s}^{-1}\;{\rm cm}^{-2}
        \else   ~photon s$^{-1}$ cm$^{-2}$\fi}
\newcommand\efluxA{\ifmmode {\rm~erg\ s}^{-1}\;{\rm cm}^{-2}\;{\rm
        \AA}^{-1} \else ~erg s$^{-1}$ cm$^{-2}$ \AA$^{-1}$\fi}
\newcommand\efluxHz{\ifmmode {\rm~erg\ s}^{-1}\;{\rm cm}^{-2}\;{\rm
        Hz}^{-1} \else ~erg s$^{-1}$ cm$^{-2}$ Hz$^{-1}$\fi}
\newcommand\cc{\ifmmode {\rm~cm}^{-3} \else cm$^{-3}$\fi}
\newcommand\FWHM{\ifmmode {\rm~FWHM} \else ${\rm~FWHM}$\fi}
\newcommand\Zsun{\ifmmode Z_{\odot} \else $M_{\odot}$\fi}
\newcommand\Lsun{\ifmmode L_{\odot} \else $L_{\odot}$\fi}

\newcommand\hbeta{\ifmmode {\rm H}\beta \else H$\beta$\fi}
\newcommand\Kalpha{\ifmmode {\rm K}\alpha \else K$\alpha$\fi}
\newcommand\nh{\ifmmode N_{\rm H} \else N$_{\rm H}$\fi}

\newcommand{\lum}{erg\,s$^{-1}$}

\newcommand{\mnras}{MNRAS}

\newcommand{\mac}{\rm~MACS J0553.4-3342}

\title[MACS J0553.4-3342]{MACS J0553.4-3342: A young merging galaxy cluster caught through the eyes of Chandra and HST}
\author[Pandge et al.] {M. B. Pandge$^{1}\thanks{E-mail:
mbpandge@gmail.com}$, Joydeep Bagchi$^{2}$,  S. S. Sonkamble$^{3}$, Viral Parekh$^{4}$, M.K. Patil,$^{3}$ \newauthor Pratik Dabhade$^{2}$, Nilam R. Navale$^{1}$, Somak Raychaudhury$^{2,5}$, Joe Jacob$^{6}$ \\
$^{1}$DST INSPIRE Faculty, Dayanand Science College, Barshi Road, Latur 413512, Maharashtra, India\\
$^{2}$Inter-University Centre for Astronomy and Astrophysics, Post Bag 4, Ganeshkhind,  Pun\'e 411007, India \\
  $^{3}$School of Physical Sciences, Swami Ramanand Teerth Marathwada University, Nanded 431606, India\\
  $^{4}$Raman Research Institute, C. V. Raman Avenue, Sadashivnagar, Bengaluru 560080, India\\
  $^{5}$Department of Physics, Presidency University, 86/1 College Street, Kolkata 700073, India\\
$^{6}$Newman College, Thodupuzha, Kerala, 685584, India. 
}

\begin{document}
\pagerange{\pageref{firstpage}--\pageref{lastpage}} \pubyear{2017}
\maketitle
\begin{abstract}
  We present a detailed analysis of a young merging galaxy cluster \mac~(z=0.43),
  from {\it Chandra} X-ray and {\it Hubble Space Telescope}
  archival data. X-ray observations confirm
  that the X-ray emitting intra-cluster medium (ICM) in this system
  is among the hottest 
  (average $T=12.1 \pm 0.6$ keV) and most luminous known.
  Comparison of X-ray and optical images confirm that this system hosts two 
  merging subclusters SC1 and SC2, separated by a projected
  distance of about 650\,kpc. The subcluster SC2 is newly identified
  in this work, while another subcluster (SC0), previously thought to be  part of this merging system, is shown to be possibly a foreground object.
  Apart from two subclusters, we
find a tail-like structure in the X-ray image, extending to a projected distance of $\sim$1\,Mpc, along the north-east direction of the eastern subcluster (SC1). From a surface brightness analysis, we detect two sharp surface
brightness edges at $\sim$40$\arcsec$ ($\sim$320\,kpc) and
$\sim$80$\arcsec$ ($\sim$640\,kpc) to the east of SC1. The inner
edge appears to be associated with a merger-driven cold front, while the outer one
is likely to be due to a shock front, the presence of which,
ahead of the cold front,
makes this dynamically disturbed cluster interesting. 
Nearly all the early-type galaxies belonging to the two subclusters,
including their BCGs, are part of a well-defined red sequence. 
\end{abstract}

\begin{keywords}
galaxies:active-galaxies:general-galaxies:clusters:individual:MACS J0553.4-3342-inter-cluster medium-X-rays:galaxies:clusters
\end{keywords}
\section[1]{Introduction}

Galaxy clusters are among the most massive gravitationally bound systems in
the Universe, assembled from the hierarchical merging of
smaller sub-haloes over cosmic time. Evidence of such interactions
among the smaller systems linger in present day clusters in the form
of subclustering in the distribution of galaxies, and in the hot intra-cluster medium
(ICM) in the form of cold fronts, shock heating
\citep{2007PhR...443....1M,0004-637X-716-2-1118}, turbulence, and sub-structure
\citep{2015ApJ...812..153O,2016MNRAS.458..681D,2016MNRAS.460L..84B}. Signatures
of such mergers can also be observed as diffuse 
non-thermal synchrotron radio emission in the form of
radio haloes and relics
\citep[e.g][]{2002NewA....7..249B,2006Sci...314..791B,2011ApJ...736L...8B,
  2012A&ARv..20...54F,2012MNRAS.426...40B}. Cluster mergers provide
ideal settings for detailed studies to understand important aspects
of the physical processes involved in these mergers, including the
thermodynamics of the hot gas, magnetic field amplification, and
high-energy particle acceleration (cosmic rays) by shocks and
turbulence
\citep{2008ApJ...679.1173R,2009A&A...506.1083V,2012MNRAS.426...40B,2015ApJ...798...90Z},
and offsets between the gas and the dark matter (DM) subclusters.

\begin{table*}
\begin{center}
  \caption{{\it Chandra} Observation log for \mac\ }
  \begin{tabular}{llrrrrlrlr}
    \hline
    \hline
    ObsID &Observing Mode &CCDs on &Starting Date  &Total Time (ks)& Clean Time (ks)\\
    \hline
    \hline
    
    12244 &VFAINT &0,1,2,3,6  & 2011-06-23&74.06 &73.28 \\
    5813  &VFAINT &0,1,2,3,6  & 2005-01-08&9.94 & 9.86\\
    
    \hline
    \hline
    \end{tabular}	
  \label{tab1}
\end{center}
\end{table*}
\begin{figure*}
\includegraphics[width=120mm,height=120mm]{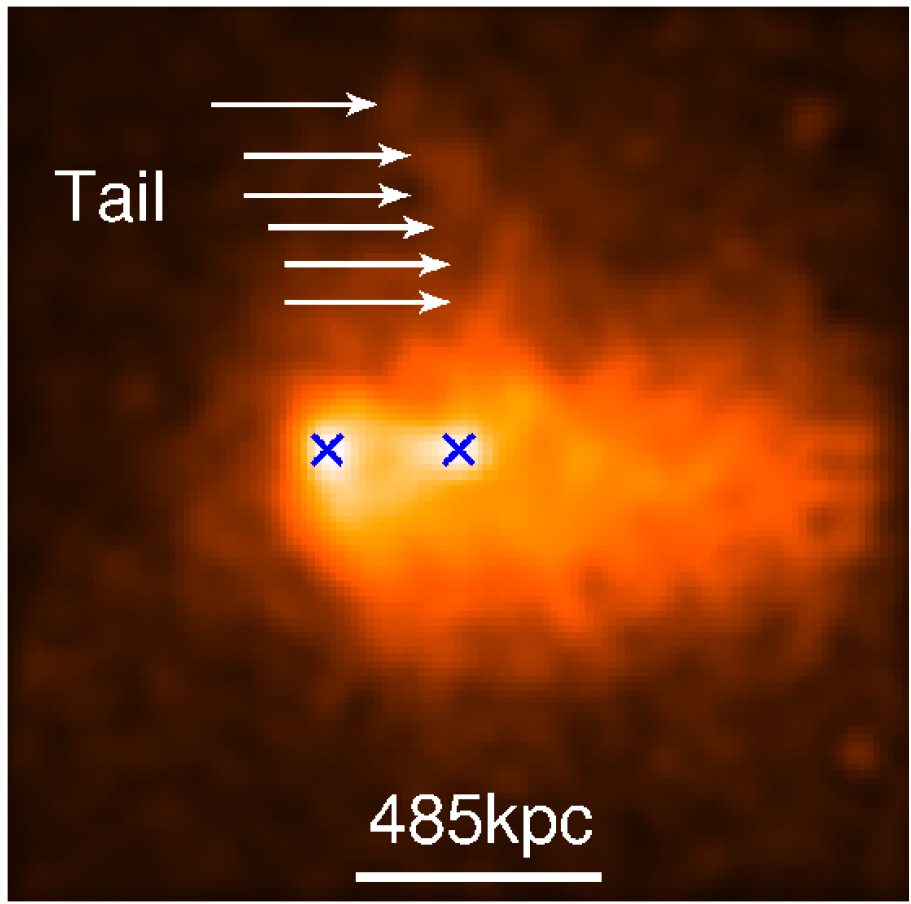}
\caption{Background subtracted, exposure-corrected central 5$\arcmin
  \times 5\arcmin$ 
         {\it Chandra} image (0.7-2.0 keV) of \mac. This image
  has been smoothed with a Gaussian kernel of width $\sigma =
  3\arcsec$. Arrows in this figure indicate the presence of an X-ray
  tail-like feature, seen along the north-east direction, appearing to
  originate from the centre of eastern subcluster (SC1). The two
  subclusters identified by \protect\cite{2012MNRAS.420.2120M} are
  highlighted by blue crosses.}
\label{fig1}
\end{figure*}
\par In this paper, we present the results from the analysis of an
83~ks {\it Chandra} X-ray observation, along with archived Hubble
Space Telescope (HST) optical observations of the extremely hot,
massive and X-ray luminous merging galaxy cluster
\mac~\citep[z=0.43,][]{2012MNRAS.420.2120M}. X-ray and radio studies
of this cluster have been reported earlier by
\cite{2012MNRAS.420.2120M} and \cite{2012MNRAS.426...40B}, using the
shallower {\it Chandra} (9.86 ks) and Giant Metrewave Radiotelescope
(GMRT) observations, respectively, where a disturbed X-ray structure
and a radio halo extending over $\sim$1.3~Mpc scale have been
reported. The joint X-ray and optical study presented in
\cite{2012MNRAS.420.2120M} pointed out that this system appears to
result from an ongoing merger of two subclusters of similar mass.
We present below a detailed morphological and thermodynamical analysis
of the distribution of the ICM in this system. This paper
also investigates the  evidence for shock and cold fronts, to
better understand the merger scenario in this system. \\

The structure of this paper is as follows. In \S2, we describe the
X-ray data reduction and imaging. \S3 represents an optical analysis, including an optical identification of the subclusters, and a colour-magnitude diagram of the subcluster. \S4 presents the
spatial and spectral analyses of the X-ray data, including surface
brightness profiles, and the spatial variation of the temperature of
the ICM in the form of a two-dimensional map. Results derived from the
present study are discussed in \S4, while \S5 describes the main
conclusions of the study. Throughout this paper, we assume
$\Lambda$CDM cosmology with $H_0$ = 73 km\, s$^{-1}$ Mpc$^{-1}$,
$\Omega_M$=0.27 \& $\Omega_{\Lambda}$=0.73, translating to a scale of
8.091\,kpc\,arcsec$^{-1}$ at the redshift z=0.43 of \mac. All spectral
analysis uncertainties are reported at the 90$\%$ confidence level,
while all other uncertainties are given at 68$\%$ confidence level.

\begin{figure*}
\center
\includegraphics[scale=1]{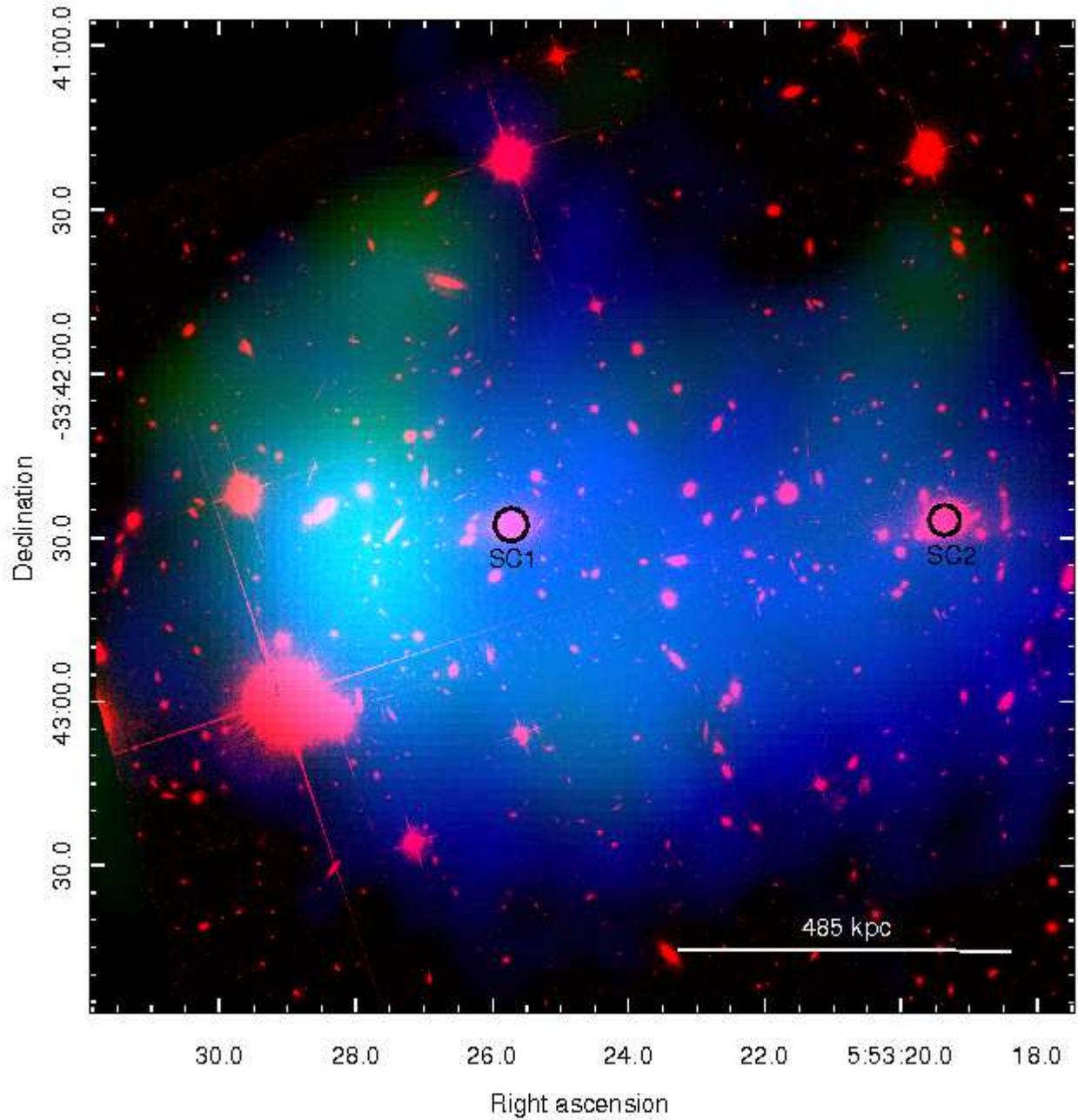}
\caption{Tri-colour image of \mac\ obtained using 0.7$-$4.0\,keV \textit{Chandra} X-ray data (shown in blue colour),  HST optical $I$ band (F814W)  data (red colour) and the GMRT 323 MHz data (green colour). This figure reveals the optical counterparts of the two subclusters (SC1 and SC2).}
\label{fig9}
\end{figure*}
\begin{figure*}
\includegraphics[scale=1]{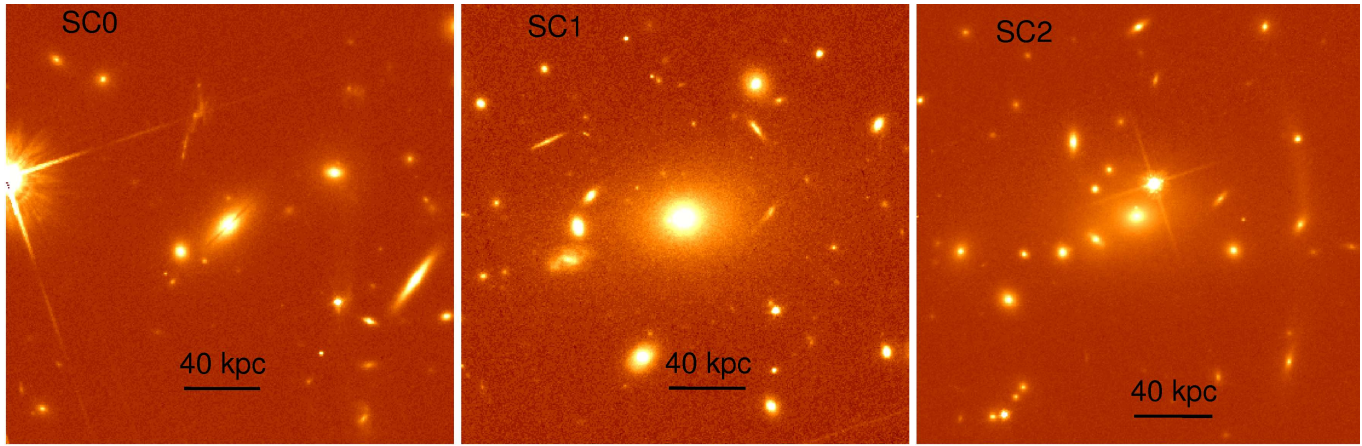}
\caption{
  HST 
  $I$-band (F814W) images of three regions of the \mac\ system,
  centred on SC0, SC1 and SC2 respectively, each 30$\arcsec \times 30\arcsec$ in size. A disk galaxy with a prominent dust lane dominates SC0.
 }
\label{optstamp}
\end{figure*}

\begin{figure*}
\center
\includegraphics[width=120mm,height=120mm]{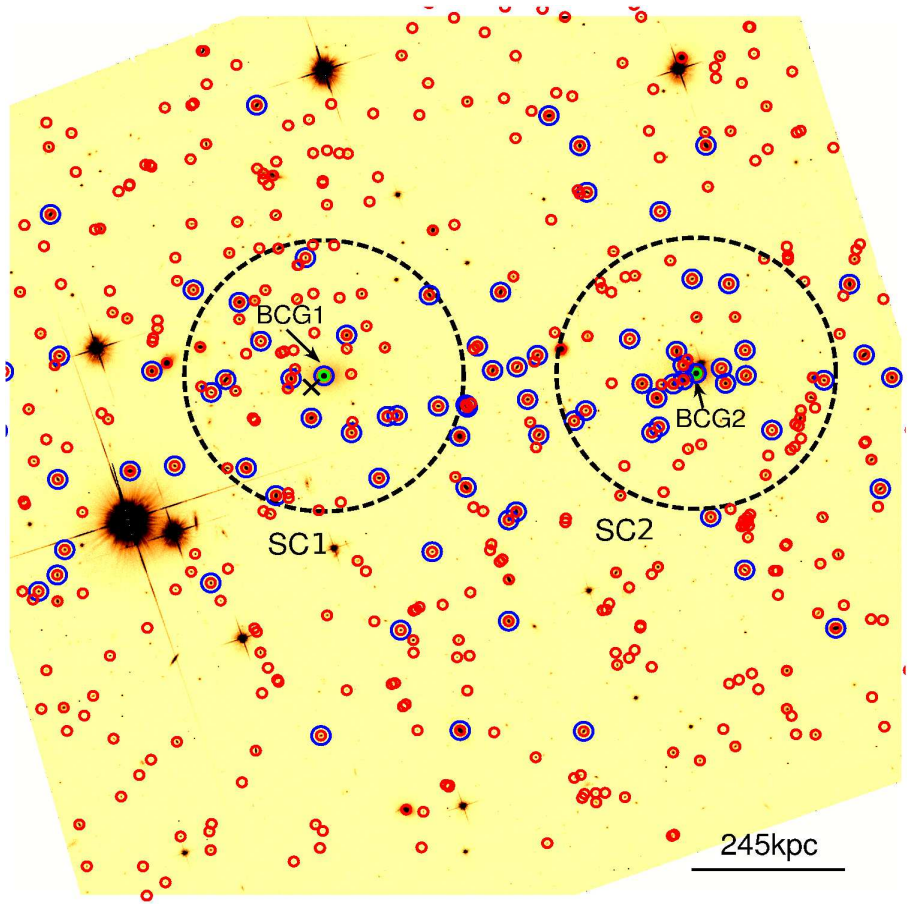}
\caption{
  An HST image of  (F814W) the central
   $\sim$ 3.3$\arcmin \times 3.3\arcmin$ region of \mac.
The subclusters SC1 and SC2 are indicated,
  and their BCGs and neighbouring members from the inner
  30$\arcsec$ are shown by black dotted circles. The red circles indicate
  the galaxies with $I$-band magnitude in the range of 18.5$-$27,
  while the blue circles mark the galaxies with $V-I$ colours in the
  range  $V-I=1.0-1.25$. The point ``x" indicates the position of
  the X-ray peak associated with the subcluster SC1.}
\label{Fig5_1}
\end{figure*}

\begin{figure*}
\center
\hbox
{
\includegraphics[width=85mm,height=85mm]{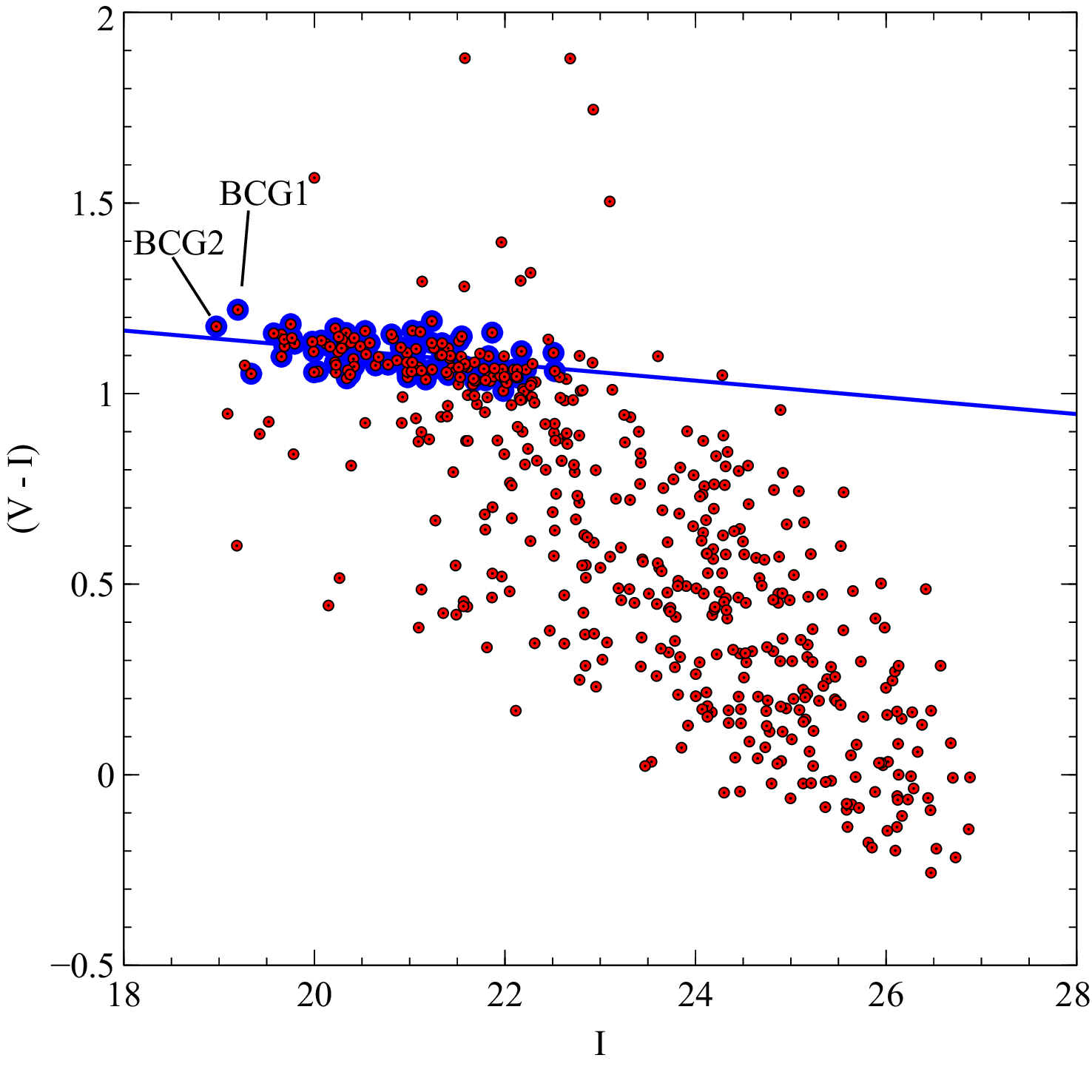}
\includegraphics[width=85mm,height=85mm]{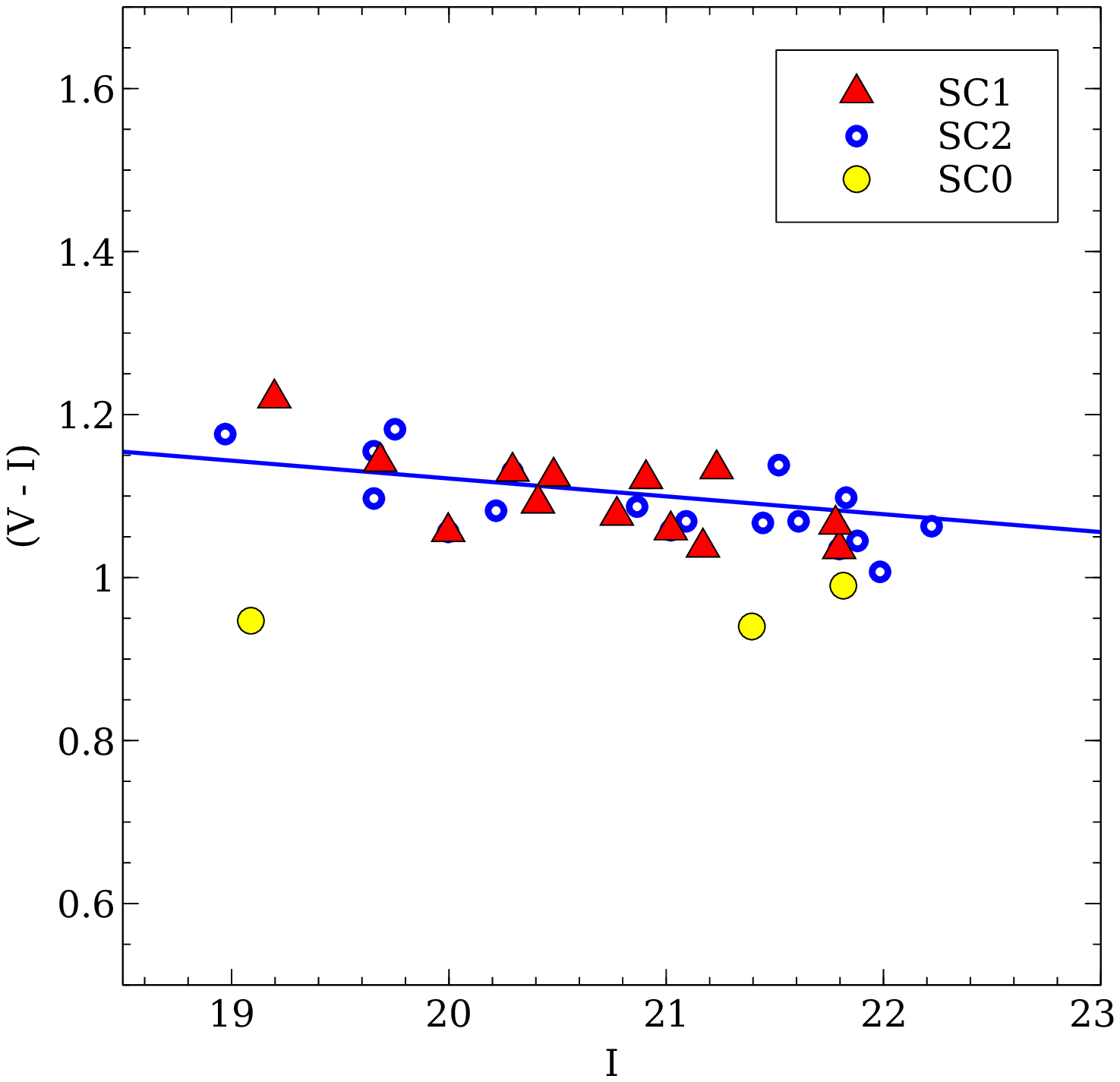}
}
\caption{{\it Left panel:} 
  {The colour-magnitude diagram for all the galaxies that have been
  identified from within the field by the Hubble extended source
  catalogue \protect\citep{2015IAUGA..2247054W}. The filled red circles show galaxies detected within the
  $\sim$ 3.3$\arcmin \times 3.3\arcmin$ field of view around the
  centre of \mac, while the blue circles are the red-sequence galaxies   within the colour range $V-I=1.0-1.25$.} {\it Right panel:} A zoomed colour-magnitude diagram of all galaxies within the central
$30\times 30 \arcsec$ of SC0 (filled yellow circles), SC1 (red
triangles) and SC2 (open blue circles).  The colour-magnitude diagram
of SC0 indicates that the prominent galaxies in this sub-cluster are
bluer in colour than the red sequence in SC1 and SC2, and therefore
possibly closer to us.}
  \label{cmdzoom}
\end{figure*}

\section[2]{X-ray data reduction and imaging}
\mac~ has been observed twice by the \chandra\ X-ray Observatory, once in January 2005 and later in June 2011, for an effective combined
exposure of 83\,ks (ObsID 5813 and 12244; for details see
Table~\ref{tab1}). Both the observations were reprocessed using the
\texttt{{CHANDRA$\textunderscore$REPRO}} task available within
CIAO\footnote{\color{blue}{{http://cxc.harvard.edu/ciao}}} 4.8,
employing the recent calibration files CALDB 4.7.2 provided by the
{\it Chandra} X-ray Center (CXC). We followed the standard \chandra~
data reduction threads\footnote{\color{blue}
  {http://cxc.harvard.edu/ciao/threads/index.html}} for the analysis
of these observations. Periods of high background flares exceeding
20\% of the mean background count rates were identified and removed
from further analysis using the {\it
  lc$\textunderscore$sigma$\textunderscore$clip} algorithm. We used
the CIAO {\tt REPROJECT$\textunderscore$OBS} task to reproject the
event files and the exposure maps in the energy range 0.7$-$7.0\,keV,
creating exposure corrected images with the {\tt
  FLUX$\textunderscore$OBS} script.

The {\tt ACIS$\textunderscore$BKGRND$\textunderscore$LOOKUP} script
within CIAO was used to identify the suitable blank sky background
fields corresponding to each of the event files. The X-ray background
files were modelled using the ``blank-sky'' datasets,
and were
reduced following the standard procedure outlined in
{\color{blue}http://cxc.harvard.edu/contrib/maxim/acisbg}. We
reprojected these sky background fields to match the coordinates of
the observations and scaled them appropriately so that their hard band
(9$-$12\,keV) count rates matched those in the science frames
before subtraction.

Point sources identified from the resultant image
using the {\ttfamily WAVDETECT} algorithm were excluded from the
subsequent analysis.  For the spectral analysis of the ICM, selected from
different regions of interest, we generated corresponding
Redistribution Matrix Files (RMF), Ancillary Response Files (ARF)
using the task {\tt SPECEXTRACT} available within CIAO. These spectra
were then exported to XSPEC \citep[version
  12.9.1,][]{1996ASPC..101...17A} for further analysis. The
exposure-corrected, background-subtracted, 0.7$-$2.0\,keV \chandra~
image of the central 5$\arcmin \times 5\arcmin$ region of \mac~is
shown in Fig.~\ref{fig1}. The blue crosses in this figure indicate the
positions of the two  previously identified subclusters (SC0 and SC1)
\citep{2012MNRAS.420.2120M}.

\section{OPTICAL ANALYSIS}

  \subsection{Optical identification of the subclusters}
To investigate the nature of optical counterparts of the
two possible  subclusters SC0 and SC1 \citep{2012MNRAS.426...40B},
 associated  with  the  two  peaks of X-ray  emission   (shown 
by the blue crosses in Fig.~\ref{fig1}), we used the three broad band
imaging observations of \mac, taken in  filters F435W ($B$),
F606W ($V$) and F814W ($I$), with effective exposure times of 4572~sec,
2092~sec and 4452~sec respectively,  from the HST archives.
Among these, we use the F814W image for
finding optical counterparts of the  possible subclusters.
We created a tri-colour map by combining  the optical F814W (shown in red), GMRT 323
MHz radio (in green) \citep{2012MNRAS.426...40B} and
0.7$-$4.0\kev~\textit{Chandra} X-ray image (in blue) observations.
The resultant  composite image is shown in Fig.~\ref{fig9}.

In Fig.~\ref{optstamp}, the HST images of
  central 30$\arcsec \times 30\arcsec$ $\sim$(250 $\times$ 250\,kpc)
  regions centred  around the brightest galaxies of  possible subclusters 
  SC0, SC1 and SC2 are shown from left to right.
In these images, we find a compact subcluster of galaxies, dominated by
a BCG, in the heart of the X-ray halo SC1 at the centre of the system
\citep{2012MNRAS.426...40B}. In addition, 
we find another subcluster,
$\sim$ 650\,kpc away to the west of SC1, marked as SC2 in this
figure, also  falling within the diffuse X-ray emission, but not showing  any bright X-ray peak
(this subcluster  is not  mentioned  in \cite{2012MNRAS.426...40B}).
In Fig.~\ref{Fig5_1}, we show the HST F814W image of an extended
$3.3\arcmin \times 3.3\arcmin$ region, where all galaxies with
$18.5<I<27.0$ mag are marked with a red circle, enclosed by a
further blue circle if their colour $V-I$ is in the range 1.0--1.25. The inner $30\arcsec$ of the subclusters SC1 and SC2 are also indicated.


\begin{figure*}
\includegraphics[width=80mm,height=75mm]{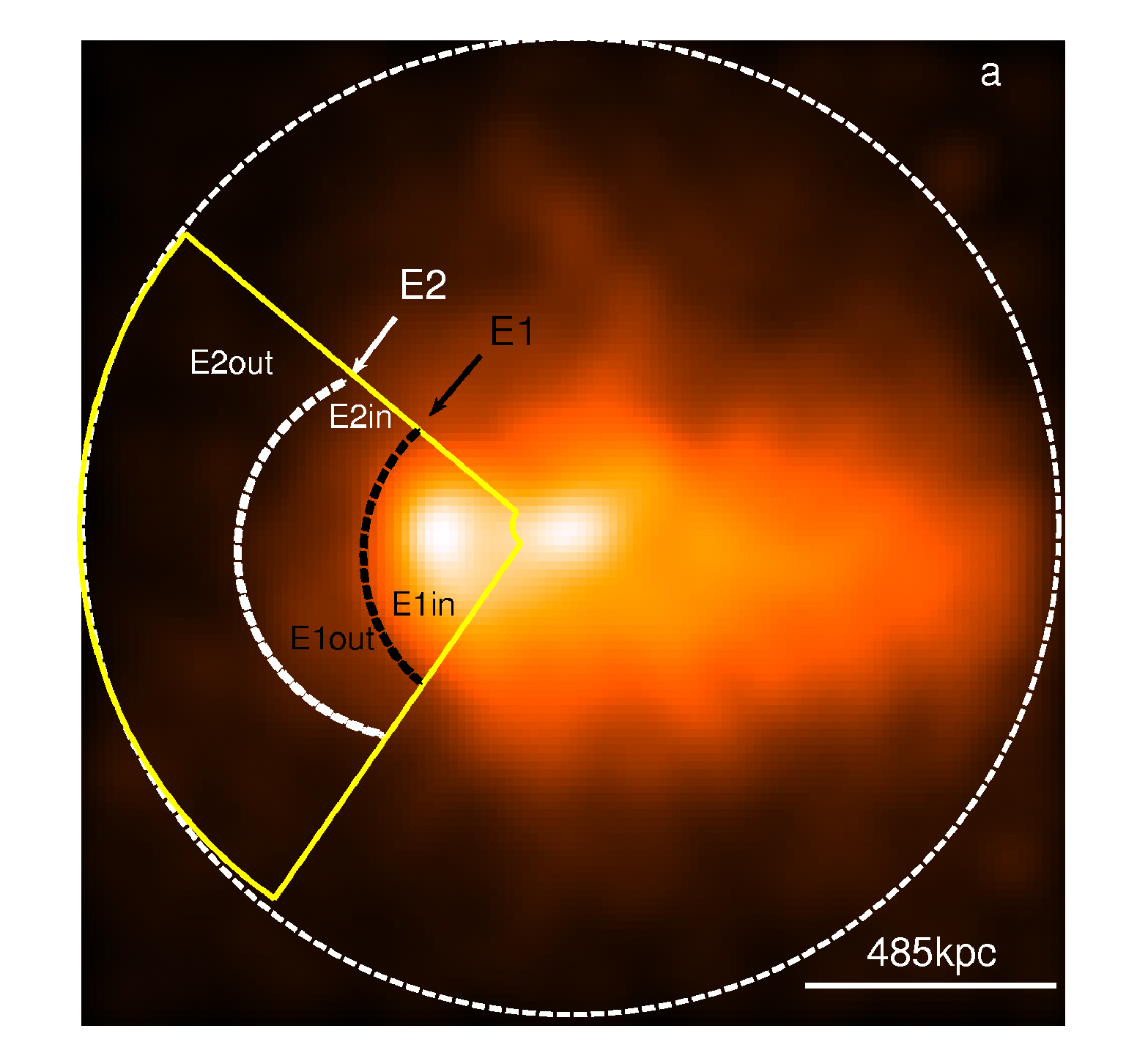}
\includegraphics[width=80mm,height=75mm]{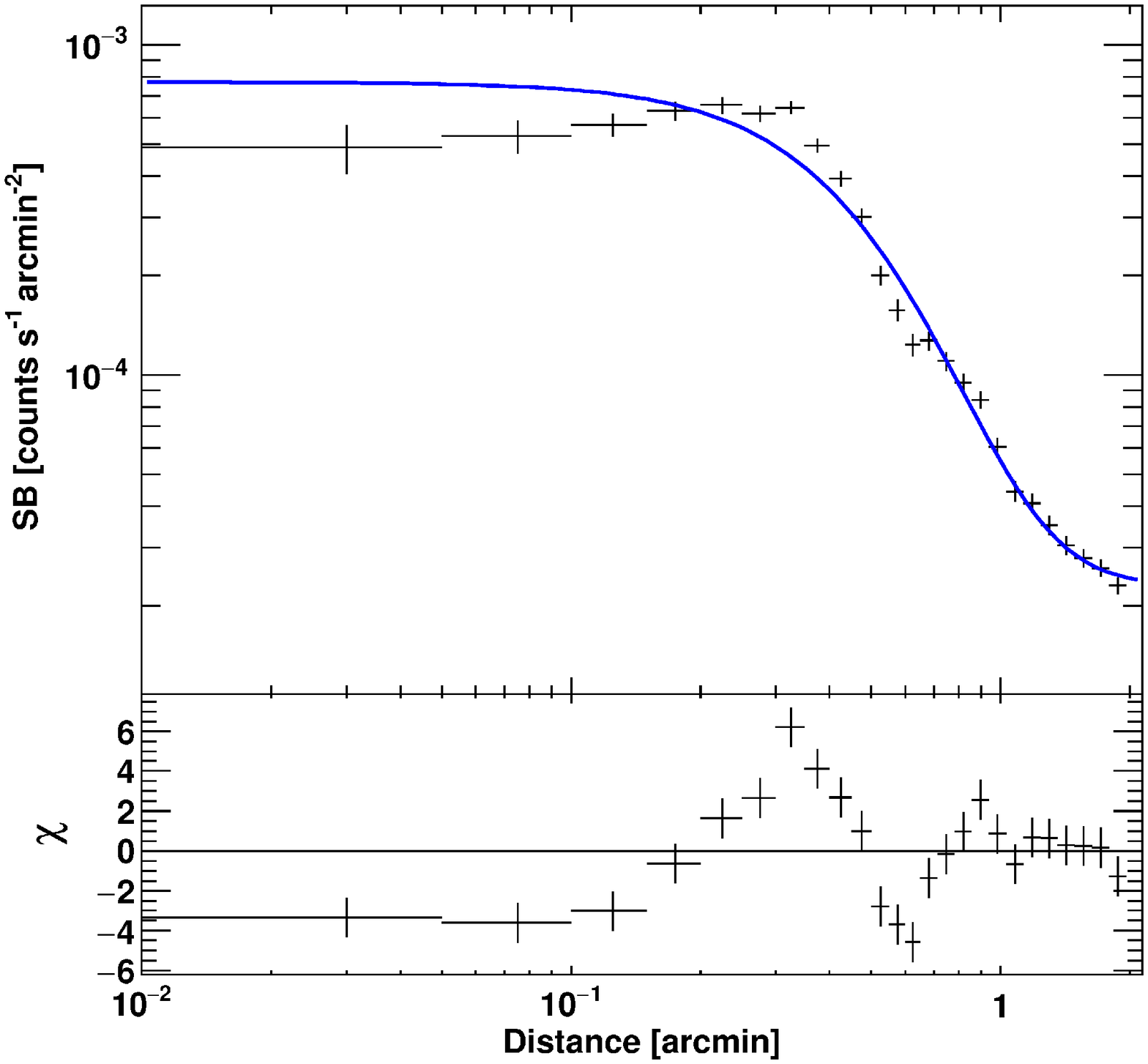}
\caption{{\it left panel:}  The 0.7$-$4.0 keV {\it  Chandra} image used
  for  the  extraction  of  the  surface  brightness  profile  of  the
  distribution of  the ICM within  \mac. The highlighted  wedge shaped
  arcs  are for  extracting  profiles for  the  identification of  the
  discontinuities in  the surface brightness distribution.  {\it right
    panel:} Projected  radial surface  brightness distribution  in the
  energy range  0.7$-$4.0\,keV of  \mac.  The  continuous line  in this
  figure indicates the best-fit 1D $\beta$-model to the data points
  (black crosses).}
\label{fig3_1}
\end{figure*}

\subsection{Colour-magnitude diagram of the subclusters}
\label{cmd}
It is well-known that the early type galaxies (ellipticals and
lenticulars) (ETGs hereafter) in clusters mostly lie in the core where
the density of galaxies is higher, while the late-type (e.g. spirals)
galaxies are predominant in the outskirts
\citep[e.g.][]{2013A&A...556C...4N,2014A&A...565A..13M}. The
colour-magnitude diagrams (CMD)
show that the early-type
members mostly appear along a well defined \textit{red sequence}
\citep[e.g.][]{2004AJ....127.2484H,2005yCat..21570001G,
  2005ApJ...634L.129S,2005ApJ...623..721P, 2005ApJ...625..121M}.
The late type galaxies scatter in such a plot with bluer colours, and
the bimodal colour-magnitude diagram provides a
useful tool to examine the properties of the galaxy population of a
massive cluster. In particular, in cases where redshifts for members are
not widely available, the CMD provides a good way of identifying ETG member
galaxies of clusters, and quantifying field contamination
\citep[e.g.][]{1998ApJ...492..461S,2001MNRAS.321...18K,2007MNRAS.374..809D}.
However, this method becomes progressively uncertain for clusters
at higher redshift.

As redshifts for each of the cluster member of \mac\ are not available
in the literature, here we construct the colour-magnitude diagram for
the \mac\ field for finding the ETG membership of SC1 and SC2, by
plotting the ($V-I$) colours of the
galaxies versus their $I$ band magnitudes from these HST observations,
within the $R_{500}$ = 1.5~Mpc of the cluster centre Fig.~\ref{cmdzoom} ({\it left panel}).
We have selected only the
extended sources in the field with flag=1 or concentration index CI $> 1.5$. This plot clearly shows the \textit{red-sequence} of early-type
galaxies in the core of \mac, with the $I$ band magnitudes in
the range between 18.5$-$23.5 and the $V-I$ colour-cut between
$\sim$1$-$1.25. This CMD shows that nearly
all the red galaxies with their $V-I$ colour values in this
range are part of the same cluster. Interestingly, the
brightest cluster galaxies (BCGs) associated with the subclusters SC1
and SC2, highlighted by black circles in Fig.~\ref{fig9},
appear close to one another confirming their membership. Their position in CMD plane is also shown in Fig.~\ref{cmdzoom} ({\it left panel}). It is also noted that nearly all the members within 30\arcsec of the BCGs (marked by dotted black circles in Fig.~\ref{Fig5_1}) strictly follow the red-sequence of the ETGs.

In Fig.~\ref{cmdzoom} ({\it right panel}), the colour-magnitude information for all the member galaxies extracted from 30\arcsec circular region is highlighted by yellow filled circles for SC0,
 red filled triangles for SC1 and open blue circles for SC2.  The ``subcluster'' 
 earlier identified as SC0,  placed to the east of SC1, (Fig.~\ref{fig1}) corresponds to
a compact group of galaxies visible in the HST image, dominated by an edge-on
disk galaxy with a prominent dust lane.  The $I$-band magnitudes of
these galaxies are between 19.09 to 21.89 and  their $V-I$ colour is in between 0.97 to 0.99 shown in Fig.~\ref{cmdzoom} ({\it right panel}).  The CMD shows clearly that these galaxies of SC0 lie significantly below the red sequence corresponding to the sublcusters SC1 and SC2, and therefore the galaxy group SC0 can not be at the same redshift as SC1 and SC2. Therefore, in the rest of this paper, we will consider this galaxy group to be a  projected foreground system, and  not part of the \mac~cluster. The bright X-ray peak near to SC0 has the high luminosity (0.5$-$10.0\,keV) of  $\sim 2.7\times10^{44}$\lum (for more detail see \S\ref{Sec:SC0}) which could not have come from such a poor group of galaxies and is likely to be a part of \mac.

\section{X-RAY SPATIAL AND SPECTRAL ANALYSIS}

\subsection{Surface Brightness Profiles}
X-ray surface brightness profiles are crucial ingredients for the investigation
of shocks and cold fronts, as indicators of the merging processes
occurring on the scale of galaxy clusters
\citep{2015ApJ...812..153O,2016MNRAS.458..681D,2016MNRAS.460L..84B}.
To identify such features in the environment of \mac\ we have derived
azimuthally averaged surface brightness profiles of the X-ray emitting
gas distribution in this cluster, by extracting X-ray counts from
within the circular annuli, with their centres as indicated in
Fig.~\ref{fig3_1} ({\it left panel}). The extracted surface brightness
profile was then fitted with the one-dimensional $\beta$-model
following the $\chi^2$ statistics of \cite{1986ApJ...303..336G}, 
\begin{equation}
 \hspace{20mm} \Sigma(r)=\Sigma_0\left[ 1+\left( \frac{r}{r_0} \right)^2\right] ^{-3\beta+0.5},
\end{equation}
where $\Sigma(r)$ represents the X-ray flux at the projected distance
$r$, $\Sigma_0$ the central surface brightness, $r_0$ the core radius
and $\beta$ the slope parameter of the profile. The best fit 1D
$\beta$ surface brightness profile is shown by the continuous line in
Fig.~\ref{fig3_1} ({\it right panel}) with the best fit parameters
$\beta$ and $r_{0}$ being 0.78 and 304\,kpc, respectively.

Unlike in the case of the cool core clusters
\citep{2013Ap&SS.345..183P,2015Ap&SS.359...61S,2017MNRAS.466.2054V},
the data points in the central region of this cluster lie below the
best-fit model. A comparison of the $\beta$-model and the data points
reveal an edge or discontinuity at a radius of
$\sim$40\arcsec. Another probable discontinuity is seen at
$\sim$80\arcsec.

\subsection{Global X-ray emission}
\label{GXP}
\begin{figure}
\includegraphics[width=80mm,height=80mm]{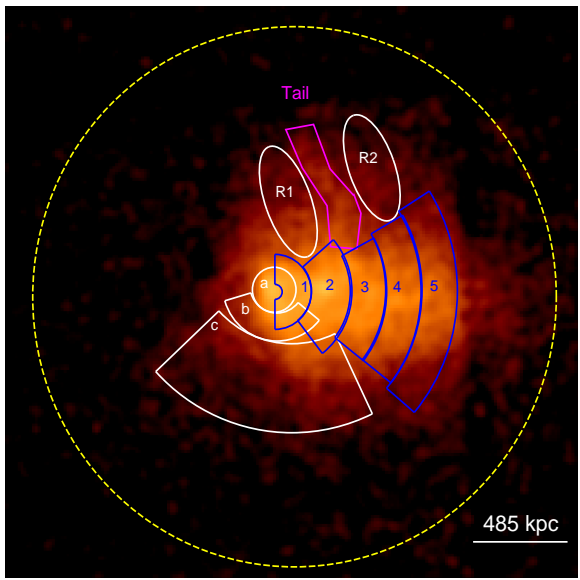}
\caption{ The {\it Chandra} image of \mac\ (energy range of
  0.7$-$7.0\,keV), delineating different regions of interest used for
  spectral extraction. The image has been exposure-corrected,
  background-subtracted, and smoothed with a 3$\sigma$-wide Gaussian
  after the removal of point sources.}
\label{fig5}
\end{figure}

To examine the global X-ray emission characteristics of the ICM in the
environment of \mac, we have extracted a cumulative 0.7$-$8.0 keV
spectrum, from a circular region of radius $R_{500}$=3.$\arcmin$09
($\sim$1500\,kpc), as shown in Fig.~\ref{fig5} (yellow dotted
circle). A corresponding background spectrum was also extracted from
the normalized blank sky background file. The spectrum was then
grouped to have at least $\sim$25 counts per spectral bin and was
imported to {\tt XSPEC} for further fitting using $\chi^2$ statistics.
We tried to constrain the spectrum with an absorbed single
temperature plasma model {\tt APEC} \citep{2001ApJ...556L..91S}, with
the Galactic absorption fixed at $N_{H}^{Gal} =
0.32\times10^{21}{\rm~cm^{-2}}$ \citep{1990ARA&A..28..215D}, 
letting all other parameters (e.g. temperature, metallicity and
normalization) vary. The best-fit minimum gives
$\chi^2=364.15$ for $383$ degrees of freedom (dof) with the elemental
abundance 0.15$\pm$ 0.06 $Z_{\odot}$ and the ICM temperature
amounting to 12.08$\pm$0.63\,keV. This implies that \mac\ represents one of the
hottest merging clusters known to us. In the $[0.1-2.4\rm,keV]$ (ROSAT-like)
band, the X-ray luminosity within the $R_{500}$ region equals
$L_{500,[0.1-2.4\rm,keV]}$ =$1.02\pm0.03\times10^{45} {~\rm
  erg~s^{-1}}$.

\subsection{Temperature map of the ICM}
\label{ICMT}
\begin{figure}
\center
\includegraphics[width=100mm,height=80mm]{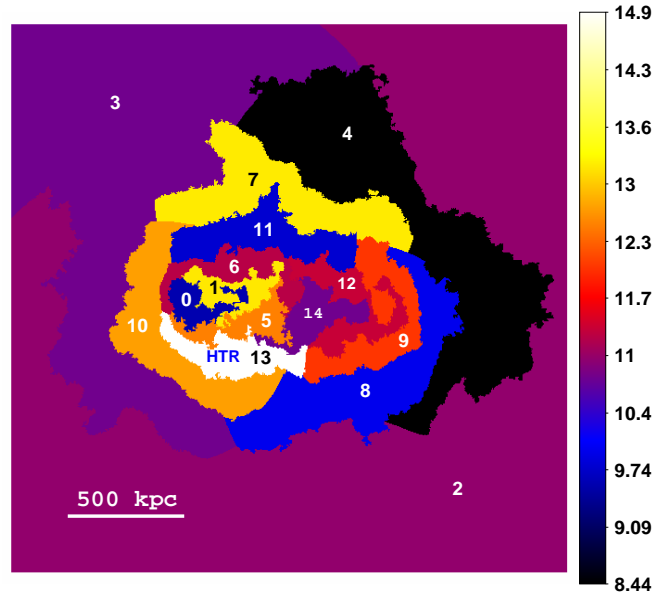}
\caption{2D temperature map of the ICM distribution within the central
  5$\arcmin$ $\times$ 5$\arcmin$ region of \mac.  Temperature values
  of the gas from different regions marked in this figure are listed
  in Table~\ref{fit_results}. Note the temperature peak (shown as
  HTR) in arc 13.}
\label{fig4}
\end{figure}
\begin{figure*}
\includegraphics[width=80mm,height=80mm]{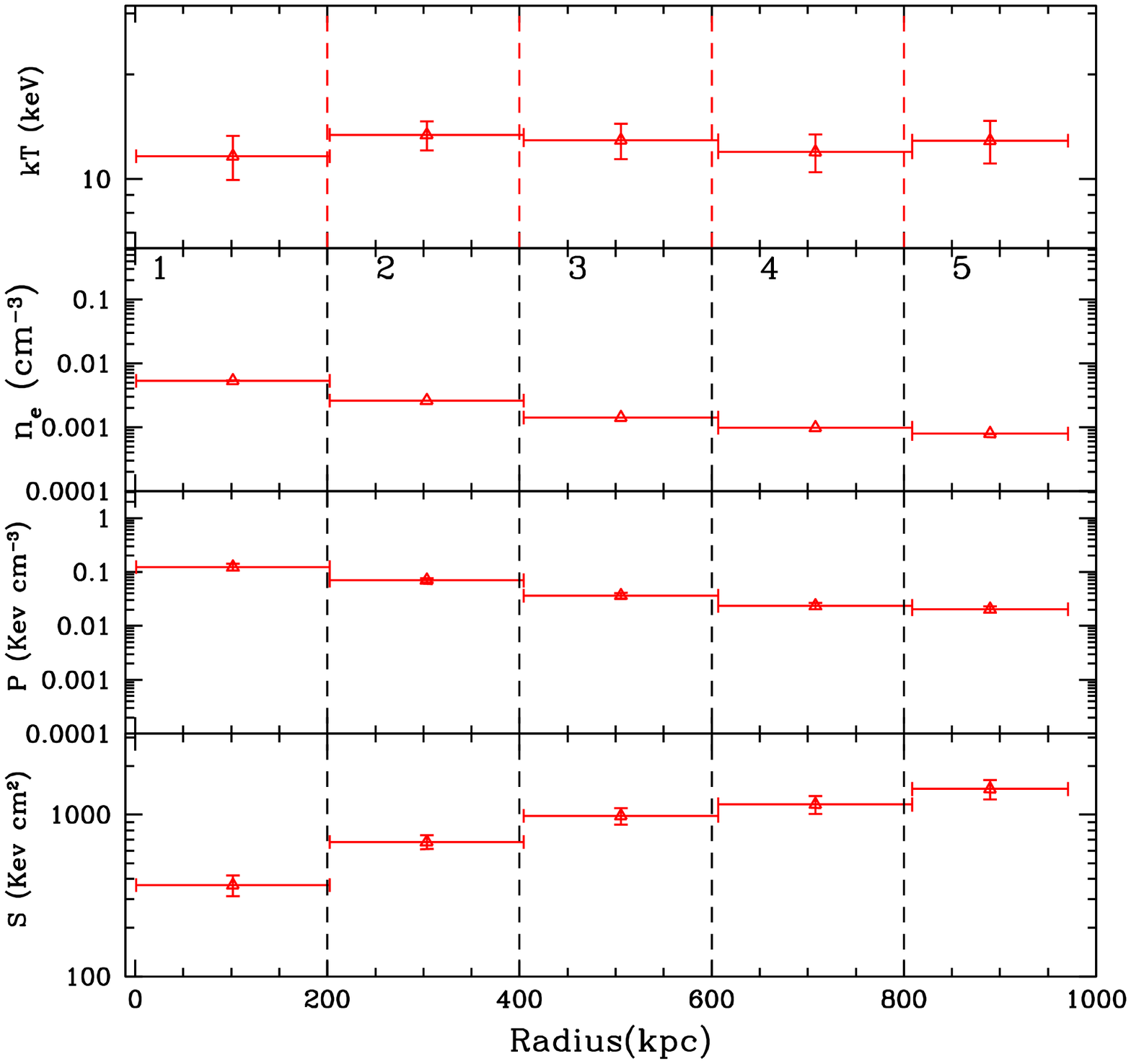}
\includegraphics[width=80mm,height=80mm]{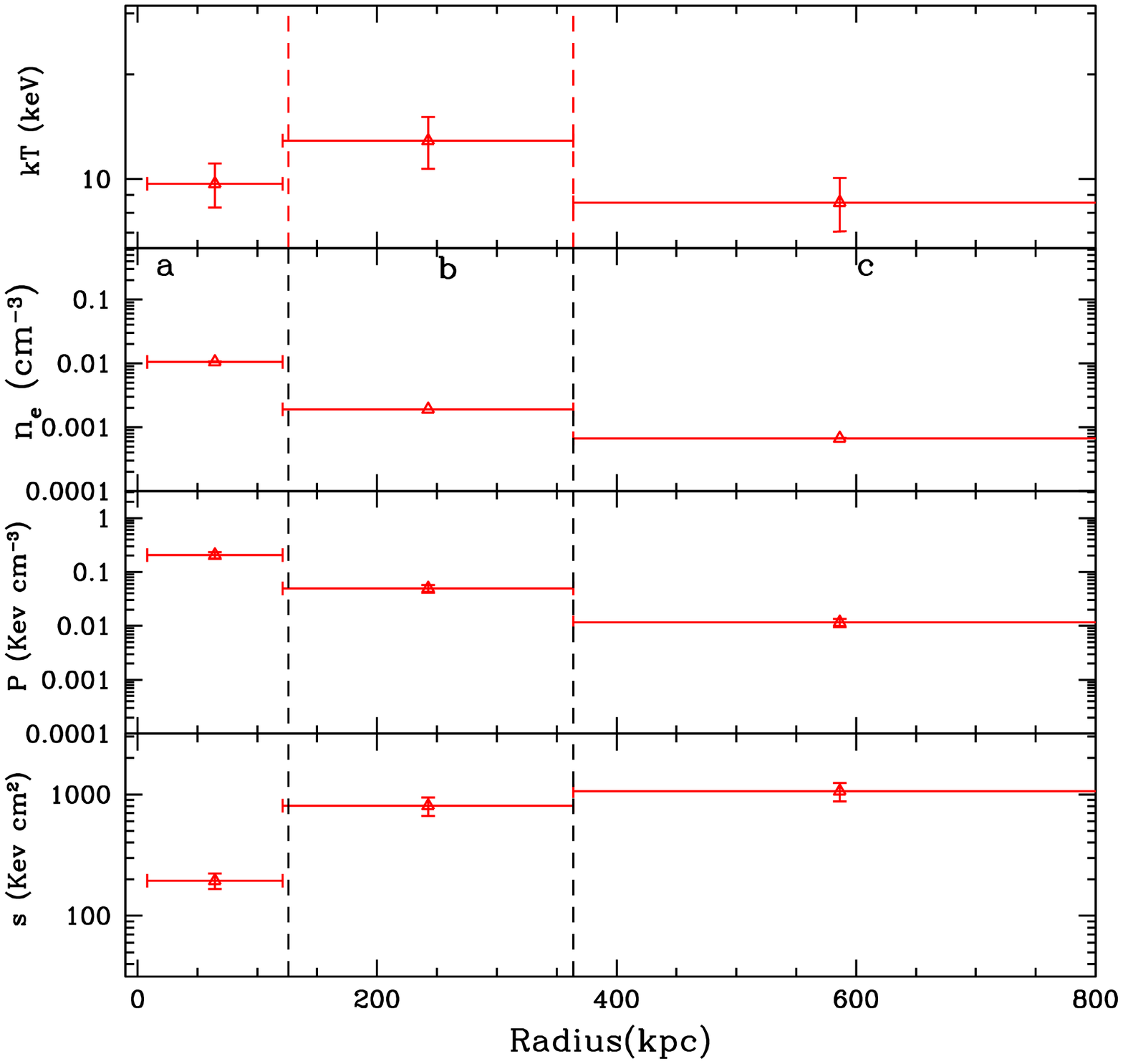}
\caption{Profiles of the thermodynamical parameters temperature (kT),
  electron density ($n_{e}$), pressure (P) and entropy (S)
  (respectively from top to bottom) for the extracted spectra from
  regions 1, 2, 3, 4, and 5 are shown in the {\it left panel}, while
  those for regions a, b and c are shown in {\it right panel}.}
\label{fig6}
\end{figure*}
We have derived a two-dimensional temperature map of the hot ICM
within \mac, following the `contour binning' technique of
\cite{2006MNRAS.371..829S}. This was achieved by generating a contour
binned image of 15 different regions, with a minimum signal to noise
ratio (S/N) $\sim$40 (i.e. 1600 counts). The regions were constrained
to the geometrical factors of 2 so that they would not be too
elongated. Spectra and response files were extracted
separately from individual bins. The spectra were then grouped to have
at least 20 counts per energy bin and were fitted with an absorbed
single temperature {\tt APEC} model as above. The best-fit temperature
values from this analysis are shown in the form of the temperature map
(Fig.~\ref{fig4}) and are also summarized in
Table~\ref{fit_results}.

This map reveals that the ICM temperature varies substantially within
the scale of the cluster, indicating its complex nature. In same figure a high temperature region (hereafter HTR, region 13) is indicated.  Another jump in the temperature of the ICM is also evident along the east of this cold front and is probably due to the presence
of a shock. Detailed properties of these cold and shock fronts are
discussed below. Notice the complexity and extended nature of the ICM
in the central region. It appears to be in homogeneously extended along the east-west direction likely due to the interactions between the two subclusters SC1 and SC2.

\par
It is possible that the cold and shock fronts exist
along the south and west directions of the X-ray centre of the
cluster. To look for them, we have extracted separate
spectra from the regions a, b and c (white semicircular arcs in
Fig.~\ref{fig5}) and regions 1, 2, 3, 4 and 5 (blue arcs in
Fig.~\ref{fig5}). The extracted spectra were treated with an absorbed
single temperature plasma code {\tt APEC} with the absorption fixed at
the Galactic value and the abundance at Z =
0.20{$\rm~$}$Z_{\odot}$. The best fit thermodynamical parameters
temperature (kT), electron density ($n_e$), pressure ($P$) and the
entropy ($S = k T\times{n_{e}^{-2/3}}$) for different regions are
shown in Fig.~\ref{fig6} and are also tabulated in
Table~\ref{fit_thermo}. The entropy, the key
parameter that records gain of the thermal energy through the shocks
and/or AGN feedback while remaining insensitive to the adiabatic
compressions and expansions, exhibits a significant increase, while
moving from region 1 through 5 (Fig.~\ref{fig6} \textit{left
  panel}). Similar rise in the entropy is also evident in the
regions a, b and c (Fig.~\ref{fig6} {\it right panel}). This
analysis failed to detect any compression due to the presence of
shocks and fronts. Similar results were also found in the surface
brightness analysis along these regions.  \par

\begin{table}
\caption{Best fit spectral properties of the ICM extracted from 15 different regions of the 2D temperature map (Fig.~\ref{fig4}).}
\begin{center}
\begin{tabular}{crcccc}
\hline
\hline
Reg.    &Net    & \(\chi^2\) (d.o.f.) & ~\rm kT   &  Norm ($10^{-4}$) \\
        &Counts             &                     & \(\mbox{\rm(keV)}\) &                 ($\rm cm^{-3}$) \\
\hline
0 & 1940  &  79.01 (77)   & $9.57\pm1.25$       & $3.75\pm0.14$ \\
1 & 1980  &  86.23 (80)   & $13.18\pm1.98$	& $3.80\pm0.10$ \\
2 & 4866  &  122.92 (128) & $11.0\pm1.86$	& $3.94\pm0.13$ \\
3 & 3055  &  103.73 (120) & $10.80\pm1.69$	& $3.30\pm0.18$ \\
4 & 2354  &  100.99 (94)  & $8.44\pm0.23$	& $6.91\pm0.36$ \\
5 & 1961  &  73.65 (76)   & $12.51\pm1.83$	& $3.56\pm0.09$ \\
6 & 1990  &  66.79 (78)   & $11.24\pm1.49$	& $3.63\pm0.14$ \\
7 & 1988  &  70.59 (79)   & $13.19\pm2.36$ 	& $3.27\pm0.09$ \\
8 & 1987  &  74.75 (80)   & $9.96\pm1.39$	& $4.16\pm0.11$ \\
9 & 2083  &  92.20 (83)   & $12.04\pm1.80$	& $3.62\pm0.17$ \\
10 & 1979  &  76.50 (80)  & $12.71\pm2.14$      & $3.45\pm0.10$ \\
11 & 1957  &  63.37 (80)  & $9.78\pm1.72$ 	& $3.43\pm0.13$ \\
12 & 1960  &  102.52 (78) & $11.35\pm1.30$	& $3.95\pm0.19$ \\
13 & 1945  &  72.41 (78)  & $14.94\pm2.13$	& $3.45\pm0.17$ \\
14 & 1936  &  68.00 (76)  & $10.78\pm1.70$	& $3.41\pm0.14$ \\
\hline
\hline
\label{fit_results}
\end{tabular}
\end{center}
\end{table}
\begin{figure*}
\includegraphics[width=85mm,height=80mm]{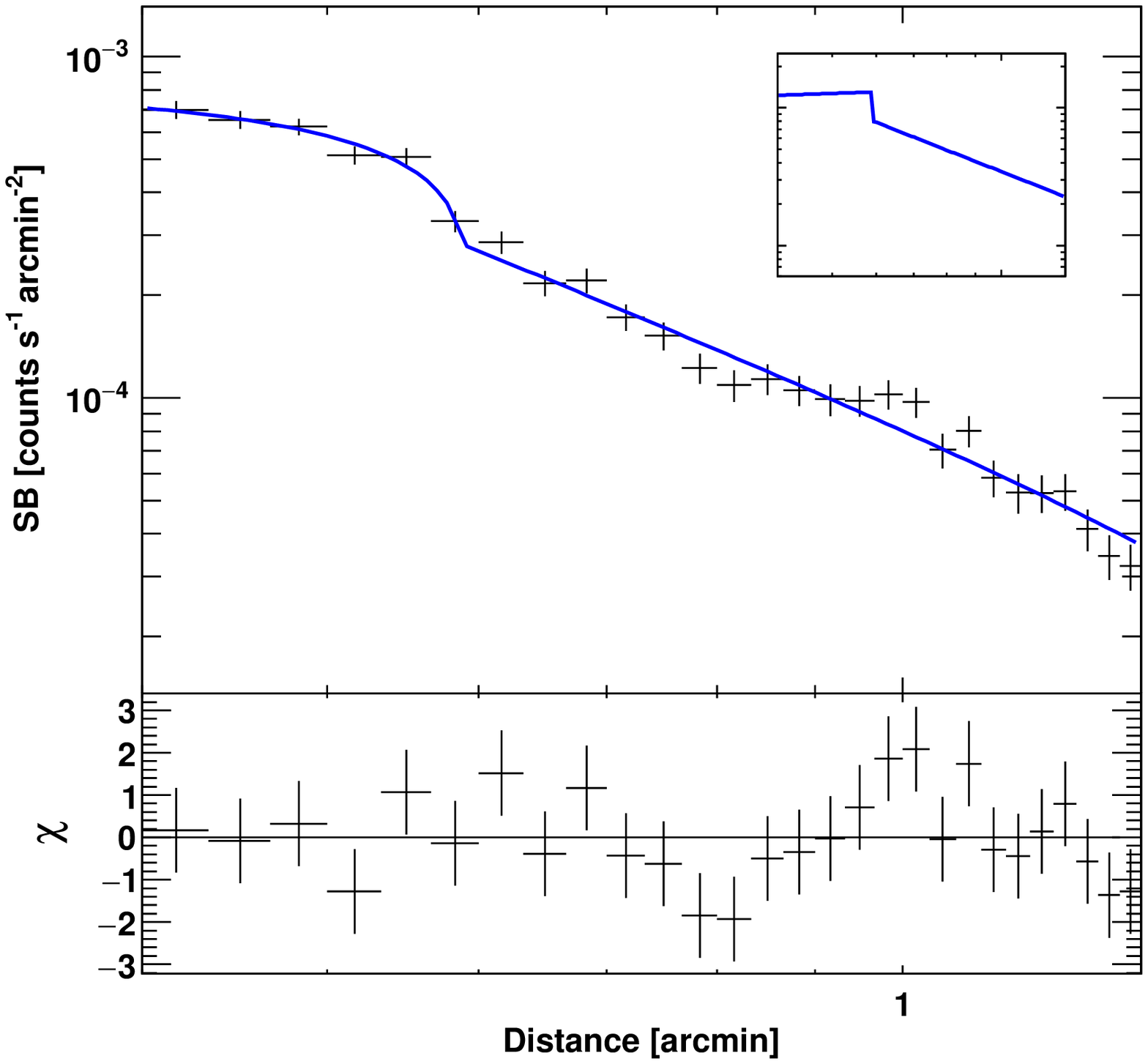}
\includegraphics[width=85mm,height=80mm]{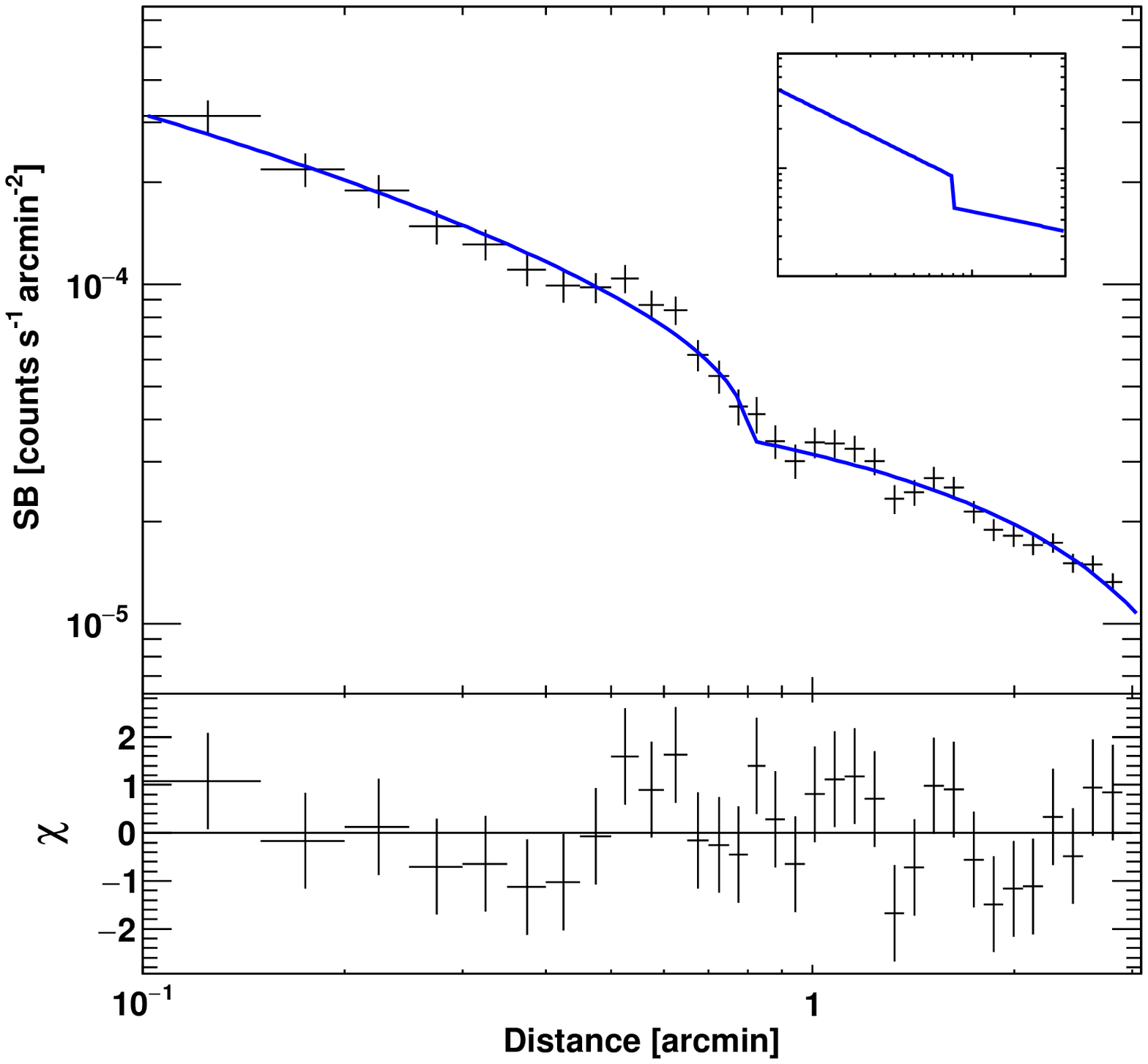}
\caption{Projected surface brightness profile extracted from the wedge
  shaped sector with opening angles between 130\degr $-$
  240\degr~around the region indicated by E1 ({\it left panel}), while
  that around the edge E2 is shown in {\it right panel}. Both these
  profiles were fitted with the deprojected broken power-law density
  model whose 3D simulations are shown in the insets. Note the jumps
  in the surface brightness near both the edges E1 and E2.}
\label{fig3}
\end{figure*}

\begin{table}
  \caption{Best fit thermodynamical parameters (temperature, electron density,
    pressure, entropy)
    of the ICM extracted from different regions in Fig.~\ref{fig5}. }
\begin{center}
\begin{tabular}{crcccc}
\hline
\hline
Reg.  & kT     &  $n_{e}$                 & P                    &  S                \\
      & ($\kev$)  &($10^{-3}\rm~cm^{-3}$)    & ($\kev\rm~cm^{-3}$)   & ($\kev\rm~cm^{2}$)  \\
\hline
1     & $11.63\pm1.20$ 		& $5.34\pm0.06$ &  $0.12\pm0.018 $  & $367\pm57$ \\
2     & $13.39\pm1.20$ 		& $2.60\pm0.01$ &  $0.069\pm0.006$  & $679\pm65$ \\
3     & $12.93\pm1.50$ 		& $1.41\pm0.01$ &  $0.036\pm0.006$  & $984\pm114$ \\
4     & $11.96\pm1.00$ 		& $0.98\pm0.01$ &  $0.023\pm0.004$  & $1156\pm145$ \\
5     & $12.90\pm1.80$ 		& $0.78\pm0.01$ &  $0.020\pm0.003$  & $1440\pm200$ \\

a     & $9.68\pm1.29$  		& $10.5\pm0.01$   &  $0.21\pm0.003 $  & $194\pm28$ \\
b     & $12.91\pm2.20$ 		& $1.19\pm0.01$  &  $0.050\pm0.022$  & $804\pm137$ \\
c     & $8.56\pm1.50$  		& $0.67\pm0.01$  &  $0.018\pm0.004$  & $1060\pm185$ \\
\hline
\hline
\label{fit_thermo}
\end{tabular}
\end{center}
\end{table}

\subsection{X-ray tail}
The {\it Chandra} image of the cluster \mac~(Fig.~\ref{fig1}) has also
revealed a prominent tail-like structure that extends in the north-east
direction of the subcluster SC1 up to a distance of about
130\arcsec($\sim$ 1002\,kpc) (at 2$\sigma$). This might be the longest
tail, originating from a stripping process, ever observed in the cluster
environment. To examine the thermal properties of the
ICM in this tail-like structure, we analysed the spectra
extracted from the long magenta polygon
and its neighbouring regions R1 and R2 (white ellipses), as shown in
Fig.~\ref{fig5}. The extracted spectra were independently fitted with an absorbed single temperature {\tt APEC} model, with the abundance
fixed at 0.2 $Z_{\odot}$. The best-fit temperature values of the gas
appearing in the tail region and its neighbouring regions R1 and R2 are tabulated in Table~\ref{tail}, and are found to be equal to
11.86$\pm$2.3 keV, 13.21$\pm$4.9 keV and 13.96$\pm$5.6 keV
respectively. The comparison of these values reveals that the gas extending in the form of the X-ray tail is similar to the regions R1 and R2 within the uncertainties, implying that it has emerged in the form of a compressed ram-pressure striped tail during the major merging process that happened in the cluster. Such a release of gas in the form of luminous tail is possible as an outcome of the merger of two equally massive subclusters \citep{2004ApJ...608..179R,2014A&A...570A.119E,2015A&A...583L...2S}.

\begin{table}
\tiny
\centering
\caption{Best fit parameters of the X-ray tail and its neighbouring regions}
\begin{tabular}{cccccccc}
\hline
Regions & Counts& {~\rm kT}               &Z(fixed)           & $L_{[0.1-2.4\rm,keV]}$ &\(\chi^2\) (d.o.f.)   \\
      &        & \(\mbox{\rm(keV)}\)  & ($Z_{\odot}$)  & ${~\rm 10^{43}erg~s^{-1}}$&         \\
\hline
Tail & 3171   & 11.86$\pm$2.3 & 	0.2       & 6.90$\pm$0.10     & 103.37 (111)  \\
R1   & 1006   & 13.21$\pm$4.9 & 	0.2       & 1.16$\pm$0.06    & 35.65 (42)  \\
R2   & 1050   & 13.96$\pm$5.6 &    	0.2       & 1.41$\pm$0.10    & 35.76 (47)  \\
\hline
\end{tabular}
\label{tail}
\end{table}

\section{Discussion}
\subsection{Surface brightness edges}

\begin{table*}
\caption{Parameters obtained from the best fit broken power-law density model}
\centering
\small
\begin{tabular}{lllllllllll}\hline
Regions &$\alpha_1$& $\alpha_2$ & $r_{sh}$ & $n_0$ & C &$\chi^{2}$/dof   & Mach No. (${\cal M}$)   \\
&&&(arcmin)&($10^{-4}~{~\rm cm^{-3}}$)& &    \\ \hline
E1 &$0.58\pm0.08$&$1.68\pm0.13$&$0.65\pm0.02$&$7.59\pm0.30$&$1.60\pm0.10$&29.17/21.00&$--$\\ 
E2 &$1.23\pm0.18$ &$0.56\pm0.05$&$1.29\pm0.04$&$0.30\pm0.04$&$1.45\pm0.16$&48.96/48&$1.33\pm0.11$\\  \hline
\end{tabular}
\label{fit}
\end{table*}

The surface brightness distribution of the ICM in Fig.~\ref{fig1} clearly
shows an edge (E1) at $\sim$ 40$\arcsec$ on the east of the X-ray
centre of the cluster \mac~ (SC1). This was also evident in the radial
surface brightness profile derived above. The radial
profile also indicated the  presence of another edge or discontinuity (E2)
at $\sim$80$\arcsec$ beyond SC1. To confirm the
presence and to examine the significance of these edges relative to
the neighbouring regions, we have extracted two separate surface
brightness profiles of the X-ray emission from the regions close to E1
and E2. The wedge shaped regions selected for this extraction have the
opening angles of 130\degr $-$ 240\degr and are shown in
Fig.~\ref{fig3_1} ({\it left panel}).

The extracted surface brightness profiles in the energy range 0.7 -
4.0 keV are shown in Fig.~\ref{fig3} ({\it left panel}) and ({\it
  right panel}), corresponding to the edges E1 and E2
respectively. These figures clearly indicate a sharp discontinuity
near the inner edge E1 (Fig.~\ref{fig3} {\it left panel}), while that
near the outer edge E2 is marginally indicative (Fig.~\ref{fig3} {\it right
  panel}). To compute the compression parameters, and hence the Mach
numbers corresponding to the ICM compression at these edges, we fitted
these profiles with deprojected broken power-law density models, using
the {\tt PROFFIT} (V 1.4) package of
\cite{2011A&A...526A..79E}. These best-fit broken power-law density
models are represented by the continuous lines in the
insets of both the figures and are parametrised as:
\begin{equation} n(r)=\left\{\begin{array}{ll} \mathcal{C}n_{0}~(\frac{r}{r_{sh}})^{-\alpha1}, & \mbox{ if }\hspace{2mm}r <r_{\rm sh}\\\\
\mathcal{}n_{0}(\frac{r}{r_{\rm sh}})^{-\alpha2} , & \mbox{if}\hspace{2mm}  r > r_{\rm sh}\end{array}\right.
\end{equation}
where $n(r)$ represents the electron number density at distance $r$, $n_0$ the density normalization, $C$ the density compression factor of the shock, $\alpha_1$ and $\alpha_2$ the power-law indices, while $r_{sh}$ represents the radius corresponding to the putative edge or cold/shock front. We allowed all the parameters to vary during the fit. The best fit parameters yielded by fitting the broken power$-$law density model are listed in Table~\ref{fit}.

According to the Rankine-Hugoniot relations \citep{1959flme.book.....L}, the density compression factor $C$ at the location of the compression is related to the Mach number ${\cal M}$ as
\begin{equation}
{\cal M}=\left[\frac{2 C}{\gamma + 1 - C(\gamma -1)}\right]^{1/2} .\mbox{ }
\label{eq:machne}
\end{equation}
\noindent Here, $\gamma$ is the adiabatic index of the gas and we
assume $\gamma=5/3$ for the present case. Thus, for determining the
Mach numbers at the location of compressions, we are required to
estimate the temperature values and hence density of the gas on either
sides of the edges. This was done by extracting separate spectra from
the inner and outer sides of the surface brightness edges, e.g. ${\rm
  E1_{in}}$ and ${\rm E1_{out}}$ for E1 and ${\rm E2_{in}}$ and ${\rm
  E2_{out}}$ for E2. All the four spectra were then fitted
independently with a single temperature {\tt APEC} model with the
redshift fixed at 0.43. The best fit temperature values of the ICM
across the edge E1 are 9.49$\pm$1.12\kev~ (T$_1$) and
15.34$\pm$2.04\kev~ (T$_2$) at ${\rm E1_{in}}$ and ${\rm E1_{out}}$,
respectively. Here, the gas on the inner side of the edge E1 appears
denser and exhibits a sharp boundary, probably due to the presence of
a merger driven cold front. The measured values of the ICM pressures
 on either side are the same within the uncertainties, thereby confirming that the edge E1
is formed by a merger driven cold front. Similarly, we also compute
the best fit temperature values of the ICM across the edge E2 and are
found to be equal to 15.34$\pm$2.04\kev~ and 8.80$\pm$1.84\kev~ for
${\rm E2_{in}}$ and ${\rm E2_{out}}$, respectively.
Then we compute the corresponding Mach numbers using the relation \citep{1959flme.book.....L}
 \begin{equation}
{\cal M}=   \frac{ \left( 8 \frac{T_2}{T_1} - 7 \right) + \left[ \left( 8 \frac{T_2}{T_1} - 7 \right)^2 + 15 \right]^{1/2} }{5}    \mbox{.}
\label{eq:machT}
\end{equation}
 
The Mach numbers ${\cal M}$ computed using Eqs. 3 and 4 at the edge E2
are 1.33$\pm$0.11 and 1.72$\pm$0.36, respectively.  These estimates
along with the measured values of the ICM temperatures collectively
indicate that this edge E2 is due to a shock front. A shock front
ahead of the merger driven cold front in this system is very similar
to those observed in the \textit{Bullet cluster}
\cite{2002ApJ...567L..27M} and the \textit{Toothbrush cluster}
\cite{2016ApJ...818..204V}, justifying this renewed interest
in this cluster.

\begin{table*}
\centering
\caption{Morphology parameters for \mac\ as discussed in \S~{\ref{morpho}}}
\label{morph_val}
\begin{tabular}{cccccccc}
\hline
Cluster &$Gini$ & $M_{20}$ & $Concentration$ &  \\
\hline
\mac~ & 0.40 $\pm$ 0.0023 & -1.09 $\pm$ 0.30 & 0.85 $\pm$ 0.37 \\
\hline
\end{tabular}
\end{table*}

\subsection{The morphological planes}
\label{morpho}

We have already seen that the cluster \mac\ represents a highly
disturbed, merging system. The {\it Chandra} image (Fig.~\ref{fig1})
confirms that the large-scale X-ray emission associated with this cluster
appears to be elongated towards the west. Further, the
HST $I$ band image reveals two close subclusters SC1 and SC2
separated by about $\sim107.5\arcsec$ ($\sim$650\,kpc), pointing
towards an ongoing merging process. Therefore, it is of great interest
to compare the dynamical state of \mac\ with other clusters that
represent different stages of their dynamical phases ranging from the
highly disturbed systems to the most relaxed ones.

For this purpose, we
have made use of the three non-parametric morphology parameters
$Gini$, $M_{20}$ and Concentration index ($C$) to characterise the
degree of disturbances in these clusters
\citep{2015A&A...575A.127P}, found to be 
useful in characterising galaxy clusters according to their level of
 dynamical disturbance. The $Gini$ coefficient parametrises the flux
distribution among the image pixels, such that for
the relaxed and cool-core clusters, where the X-ray flux is concentrated only in a
small number of image pixels, its value is closer to 1, while in
non-relaxed clusters, where the flux is more widely distributed among the image
pixels, $Gini$ takes values close to 0
\citep[e.g.][]{2004AJ....128..163L}.
The moment of light $M_{20}$ is the
normalized second order moment of relative contribution of the
brightest $20\%$ pixels \citep{2004AJ....128..163L} and is a measure
of the spatial distribution of the bright cores and subclusters in the
cluster. Typically, the value of the moment of light parameter
$M_{20}$ is found to vary in the range between $-2.5$ for the case of
relaxed clusters to $-0.7$ for most disturbed systems. The third
parameter $C$ is a measure of the concentration of the flux in the
cluster and depends on the ratio of the radii at which $80\%$ and
$20\%$ of the cluster fluxes are measured
\citep{2003ApJS..147....1C}. It takes the minimum value of 0.0 for the
most disturbed clusters.

\begin{figure*}
\includegraphics[width=60mm,height=60mm]{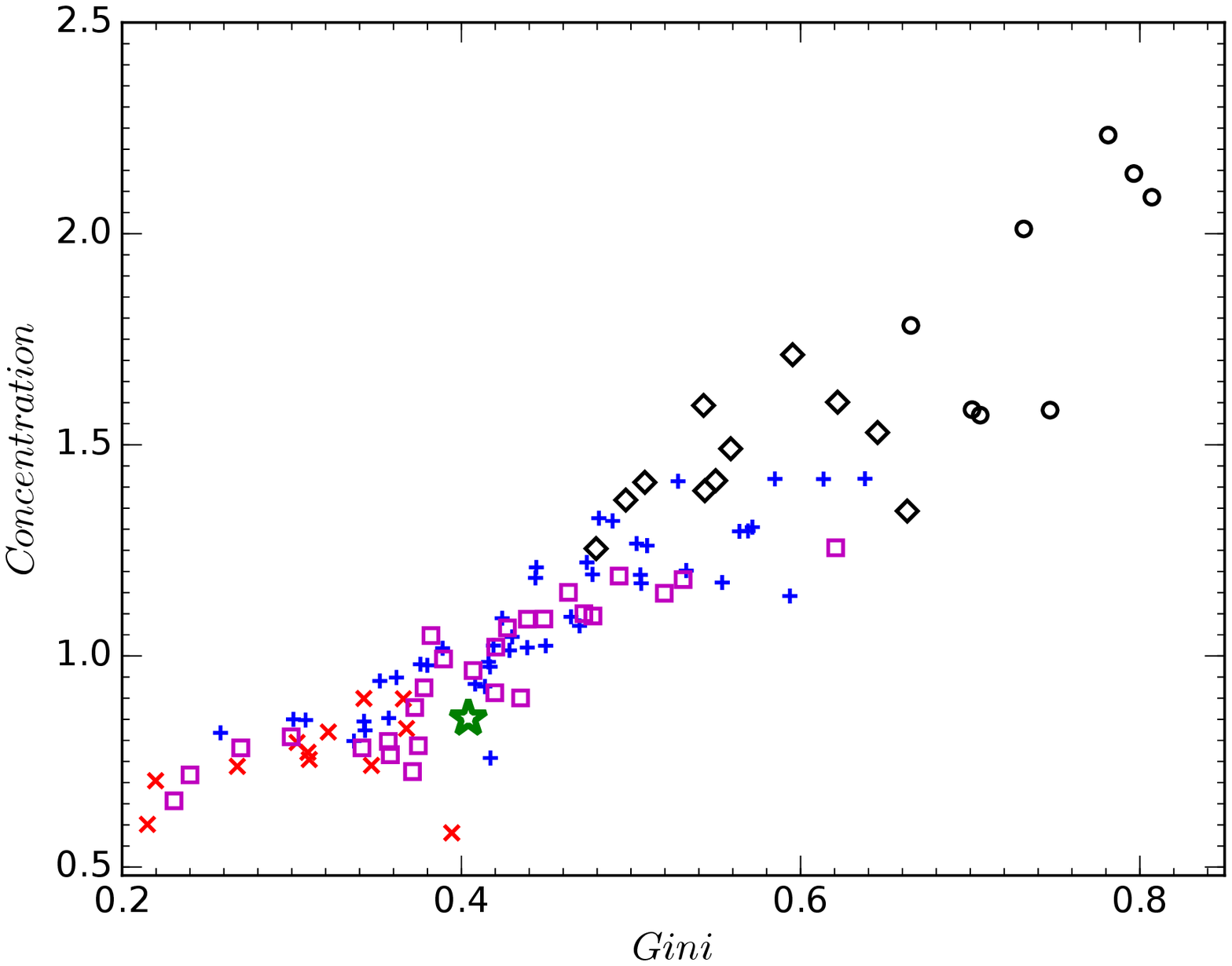}
\includegraphics[width=60mm,height=60mm]{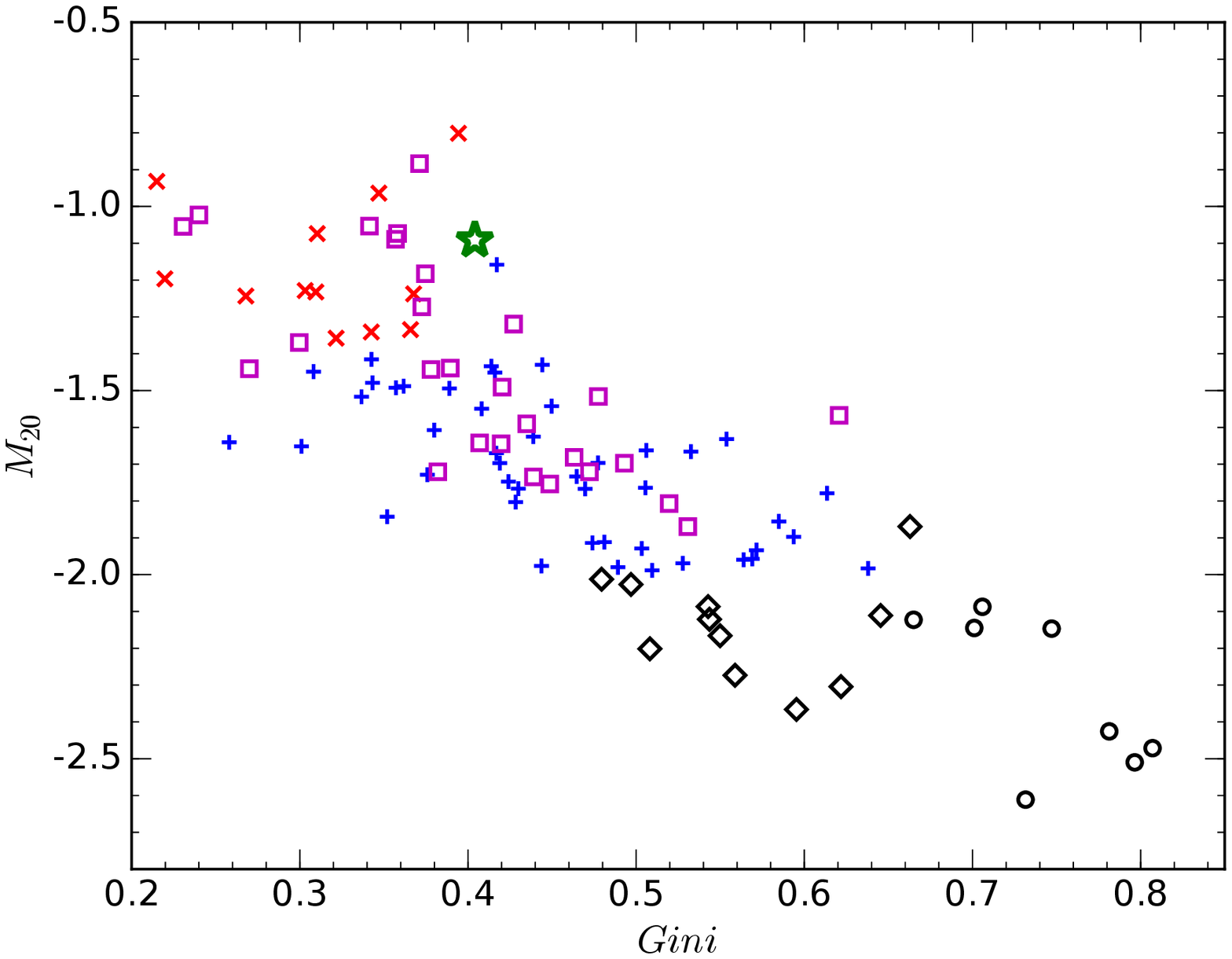}
\includegraphics[width=60mm,height=60mm]{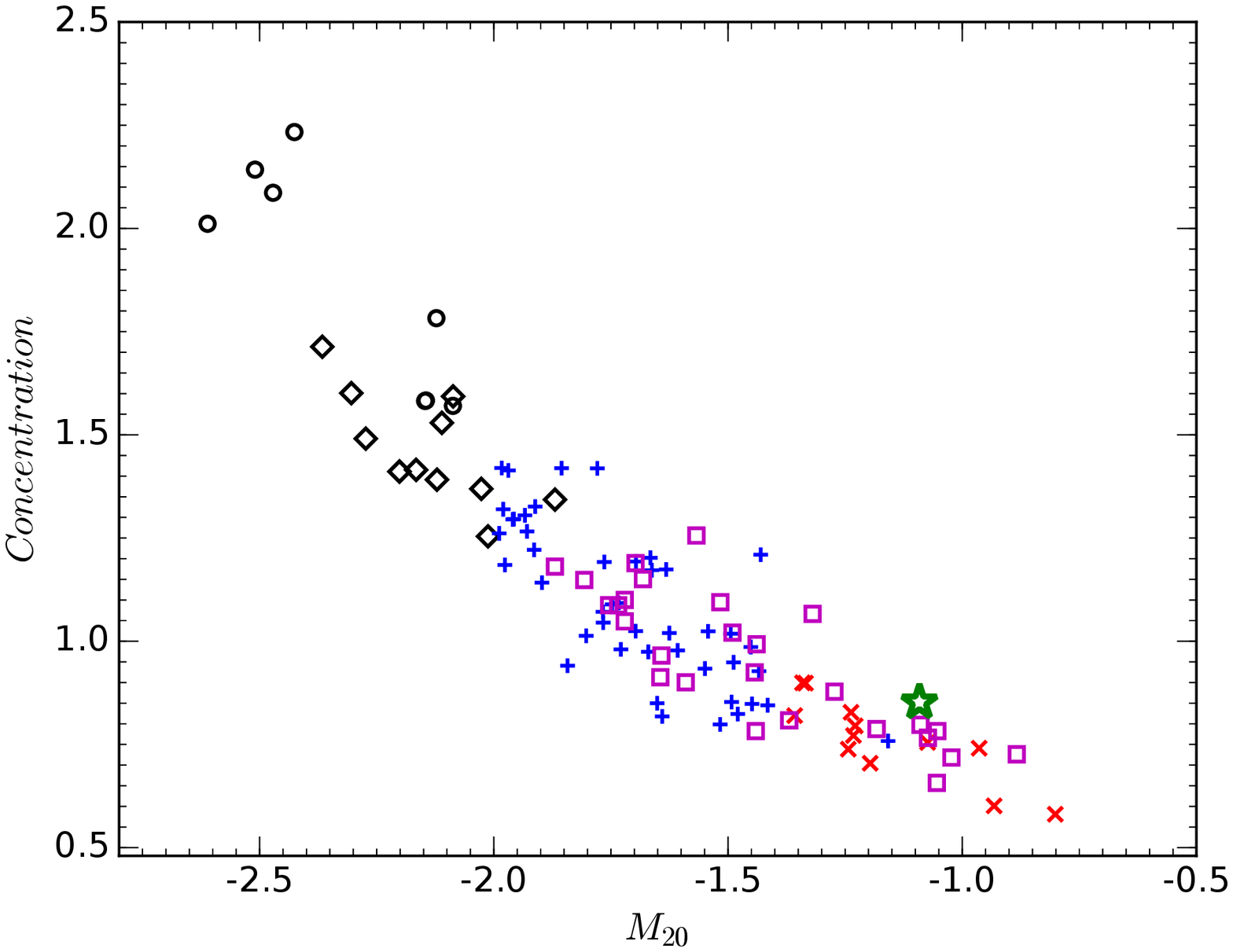}
\caption {The morphological parameter planes for the control
  sample of galaxy clusters taken from \citealt{2015A&A...575A.127P}
  for identifying their dynamical states. 
  Open circles in all the three plots
  represent the `most relaxed' clusters, diamonds the `relaxed'
  clusters, pluses the `non-relaxed', while the `most disturbed'
  clusters are indicated by crosses. The squares are clusters with
  radio halos and known to be merging clusters. The 
  position of \mac\ in these plots is indicated by a green star.
  \citep{2009A&A...507.1257G}.}
\label{fig7}
\end{figure*}
\begin{figure*}
\centering
\includegraphics[width=160mm,height=75mm]{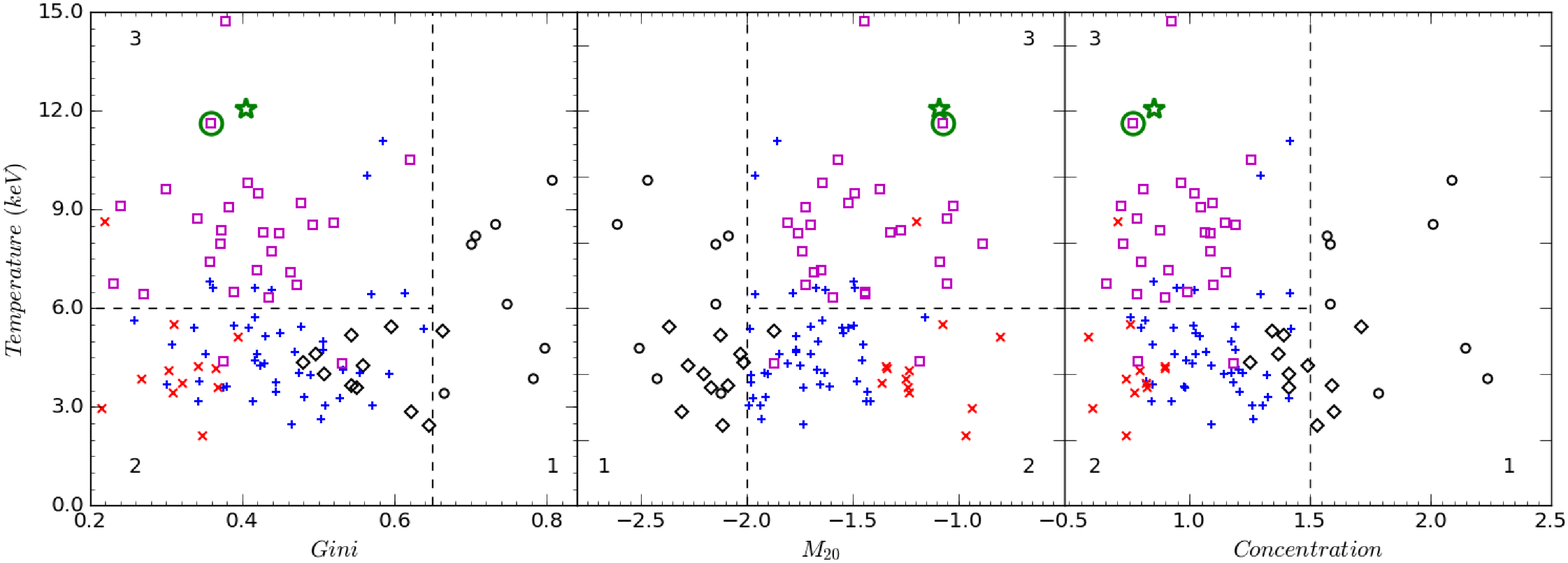}
\caption {Morphology parameters vs. temperature for the sample
  clusters as in Fig.~\ref{fig7}. We subdivided the morphology
  parameter vs temperature plot into three regions: (1) dynamically
  relaxed clusters, (2) radio-quiet (no radio halo) merger clusters,
  and (3) radio-loud (with radio halo) merger clusters. The green star
  represents \mac\ while a square within a green circle shows the `Bullet
  Cluster' 1E 0657-56.}
\label{fig8}
\end{figure*}
For estimating these morphological parameters in the case of \mac\ we
have made use of the cleaned, background and exposure-corrected
\textit{Chandra} image. The computed parameters from within the
500\,kpc region around the cluster centroid in this figure are listed
in Table~\ref{morph_val}. To compare the dynamical state of \mac~ with
those for the sample clusters of \cite{2015A&A...575A.127P}, we plot
different correlations among the morphological parameters and are
known as the \textit{morphological planes} (Fig.~\ref{fig7}). The
control sample comprises 49 low-redshift ($z = 0.2-0.3$) and 36
high-redshift ($z = 0.3 - 0.8$) clusters of different dynamical
states, representing relaxed as well as disturbed phases. Open circles
in this figure represent the most relaxed clusters, diamonds the
relaxed, pluses `+' the non-relaxed and the crosses `x' the most
disturbed systems from the sample. The squares in all three plots
represent the galaxy clusters with radio halos, known to be merging
clusters and are taken from \cite{2009A&A...507.1257G}. This figure
reveals that the the relaxed and disturbed systems takes positions on
extreme ends, while the clusters with intermediate dynamical stages
occupy positions in between them. These plots segregate clusters using
different combinations of the morphological parameters with their
limits ranging between -0.65$<$$G$$<$0.40, -1.4$<$$M_{20}$$<$-2.0, and
1.55$<$$C$$<$1.0. We show the position of \mac~ in these plots with
the green star using its morphological parameters given in
Table~\ref{morph_val}. Careful observations of all these plots reveal
that \mac~ occupies position in the family of the highly non-relaxed
clusters.

We have also plotted the morphological parameters of the control
sample including the radio-halo merging clusters of
\cite{2009A&A...507.1257G} as a function of the cluster temperature
and the resultant plots are shown in Fig.~\ref{fig8}. For better
representation we have divided these plots in three different regions:
$1.$ the systems of dynamically relaxed nature, $2.$ radio-quiet
merging clusters i.e., mergers without radio halos, and $3.$ the
radio-loud merging clusters (i.e. clusters with radio halos) with ICM
temperature T $>$ 6\,keV. In these plots we also indicate position of
\mac\ by a green star. In all the three plots \mac\ occupies
position in region~3, indicating that it belongs to the class of galaxy
clusters that are dominated by the non-relaxed dynamical state with
radio halos around them. \mac\ has also been reported to be the
hottest galaxy clusters with a 1.3 Mpc-scale radio halo
\citep{2012MNRAS.426...40B}. In these plots we also indicate,
 by a square surrounded by a green circle, 
the well-known highly disturbed, extremely hot galaxy cluster 1E 0657-56 (The "Bullet Cluster") with a radio halo
\citep{2002ApJ...567L..27M,2014MNRAS.440.2901S}. Interestingly, \mac~
appears close to this highly disturbed system in all the
three plots (Fig.~\ref{fig8}).  In view of its similarity with the
heavily disturbed systems, it is of great interest to obtain the
detailed gravitational lensing mass map of the dark-matter
distribution in this cluster and compare its distribution relative to
the baryonic and galactic components \citep{2017arXiv170603535E}. In short, our subcluster
analysis clearly demonstrates that \mac\ is a dynamically  disturbed
cluster, belonging to region 3, the region dominated by the clusters
exhibiting merging processes and radio halos around them.  \par

\subsection{The nature of the  X-ray bright peak east to the SC1}
\label{Sec:SC0}
A bright X-ray peak is apparent about $\sim$200\,kpc east of
  the eastern subcluster SC1. From our spectral analysis from X-ray
  photons extracted from a $\sim$15$\arcsec$($\sim$120\,kpc) circular
  region of this peak, fitted in the same way as discussed in
  \S\ref{GXP}, the best fit temperature and (0.5$-$10.0\,keV) band luminosity values are
  found to be 9.54$\pm$1.44\,keV and
  $\sim2.7\pm0.10\times10^{44}{~\rm erg~s^{-1}}$ (minimum
  $\chi^2=122.74$ for $96$ degrees of freedom). Comparing with the SC1
  gas temperature, it is evident that the gas temperature in this X-ray   bright peak 
  is cooler than that of SC1 (see. region 2 in
  Table~\ref{fit_thermo} $13.39\pm1.20$). This might be attributed to
  being due to the displacement of the cool dense gas from the eastern
  subcluster SC1 during the major merger event.

\section{Conclusions}

In this paper, we have presented the analysis of a total of 83\,ks of
{\it Chandra} X-ray observations, along with HST optical observations,
of \mac, one of the hottest systems known representing a merging
cluster. The main objectives of the study were to identify and confirm
the presence of different subclusters in the environment of \mac, and also
to investigate discontinuities or edges in the X-ray surface
brightness distribution that remained undetected in the previous
studies. The present study has clearly demonstrated that the ICM in
this cluster hosts two merging sub-clusters, whose merger axis lies
along the east $-$ west direction of the cluster. Important results
from this study are summarized below.

\begin{itemize}

\item Optical identification of the member galaxies in the field of
  \mac\ cluster confirms that this system actually hosts two different
  merging subclusters SC1 and SC2 separated by a projected distance of
  $\sim$650~kpc. 
  
\item The exposure corrected background subtracted image shows an
  X-ray tail-like structure extending up to a projected distance of
  130$\arcsec$ or $\sim$1002\,kpc (at 2$\sigma$ confidence) from the
  centre of SC1. The gas along this tail appears to be similar to its neighboring region within the uncertainties.

\item X-ray surface brightness profiles extracted from
  the wedge shaped regions with opening angles of 130\degr $-$
  240\degr~indicate two sharp surface brightness edges (E1 \& E2) at
  $\sim$40$\arcsec$ ($\sim$323\,kpc) and $\sim$80$\arcsec$
  ($\sim$647\,kpc) east of the centre of the cluster,
  respectively. The inner edge E1 represents a merger-driven cold front,
  while the outer edge E2 is due to a shock front. The Mach numbers
  ${\cal M}$ associated with the compression due to the shock at E2 are
  estimated to be $1.33\pm0.11$ and $1.72\pm0.36$, from a density
  compression jump analysis and from the temperature measurement on either
  sides of the shock front. A shock front ahead of the
  merger driven cold front is very similar to those seen in the
  \textit{Bullet} and the \textit{Toothbrush} clusters.

\item Spectral studies reveal that the ICM in \mac\ to be at an average
  temperature of $T_{500}$= $12.08\pm0.63\kev$, with average metallicity
  of $Z_{500}$ = 0.15$\pm 0.06$ $Z_{\odot}$\,and luminosity
  $L_{500,[0.1-2.4\kev]}$ =1.02$\pm0.03\times10^{45} {~\rm
    erg~s^{-1}}$. This makes it one of the hottest and
  brightest clusters known. 

\item The dynamical state of \mac\ is examined using the morphological
  parameters, as well as a subcluster analysis. This indicates
  that \mac\ represents a case of a dynamically disturbed
  cluster.

\item The colour-magnitude diagram plotted for \mac\ demonstrates that
  nearly all the early-type galaxies, including BCGs at SC1 and SC2,
  within 30\arcsec of the centres of the subclusters SC1 and SC2 are part of the
  same system, and lie within its well-defined red-sequence.


\end{itemize}

\section*{Acknowledgments}
MBP gratefully acknowledges the support from following funding schemes:  Department of Science and Technology (DST), New Delhi under the SERB Young Scientist Scheme (sanctioned No: SERB/YSS/2015/000534), Department of Science and Technology (DST), New Delhi under the INSPIRE faculty  Scheme (sanctioned No: DST/INSPIRE/04/2015/000108). SSS acknowledges
financial support under Minority Fellowship program, Ministry of
Minority Affairs, Government of India, (Award No
F1-17.1/2010/MANF-BUD- MAH-2111/CSA-III).  JB, PD and JJ gratefully
acknowledge generous support from the Indo-French Centre for the
Promotion of Advanced Research (Centre Franco-Indien pour la Promotion
de la Recherche Avan\'{c}ee) under programme no. 5204-2. JJ wishes to
acknowledge with thanks the support received from IUCAA, India in the
form of visiting associateship. This research has made use of the data
from {\it Chandra} Archive. Part of the reported results are based on
observations made with the NASA/ESA Hubble Space Telescope, obtained
from the Data Archive at the Space Telescope Science Institute, which
is operated by the Association of Universities for Research in
Astronomy, Inc.,under NASA contract NAS 5-26555. This research has
made use of software provided by the Chandra X-ray Center (CXC) in the
application packages CIAO, ChIPS, and Sherpa.  This research has made
use of NASA's Astrophysics Data System, and of the NASA/IPAC
Extragalactic Database (NED) which is operated by the Jet Propulsion
Laboratory, California Institute of Technology, under contract with
the National Aeronautics and Space Administration.  Facilities:
Chandra (ACIS), HST (ACS).
\def\aj{AJ}%
\def\actaa{Acta Astron.}%
\def\araa{ARA\&A}%
\def\apj{ApJ}%
\def\apjl{ApJ}%
\def\apjs{ApJS}%
\def\ao{Appl.~Opt.}%
\def\apss{Ap\&SS}
\def\aap{A\&A}%
\def\aapr{A\&A~Rev.}%
\def\aaps{A\&AS}%
\def\azh{AZh}%
\def\baas{BAAS}%
\def\bac{Bull. astr. Inst. Czechosl.}%
\def\caa{Chinese Astron. Astrophys.}%
\def\cjaa{Chinese J. Astron. Astrophys.}%
\def\icarus{Icarus}%
\def\jcap{J. Cosmology Astropart. Phys.}%
\def\jrasc{JRASC}%
\def\mnras{MNRAS}%
\def\memras{MmRAS}%
\def\na{New A}%
\def\nar{New A Rev.}%
\def\pasa{PASA}%
\def\pra{Phys.~Rev.~A}%
\def\prb{Phys.~Rev.~B}%
\def\prc{Phys.~Rev.~C}%
\def\prd{Phys.~Rev.~D}%
\def\pre{Phys.~Rev.~E}%
\def\prl{Phys.~Rev.~Lett.}%
\def\pasp{PASP}%
\def\pasj{PASJ}%
\def\qjras{QJRAS}%
\def\rmxaa{Rev. Mexicana Astron. Astrofis.}%
\def\skytel{S\&T}%
\def\solphys{Sol.~Phys.}%
\def\sovast{Soviet~Ast.}%
\def\ssr{Space~Sci.~Rev.}%
\def\zap{ZAp}%
\def\nat{Nature}%
\def\iaucirc{IAU~Circ.}%
\def\aplett{Astrophys.~Lett.}%
\def\apspr{Astrophys.~Space~Phys.~Res.}%
\def\bain{Bull.~Astron.~Inst.~Netherlands}%
\def\fcp{Fund.~Cosmic~Phys.}%
\def\gca{Geochim.~Cosmochim.~Acta}%
\def\grl{Geophys.~Res.~Lett.}%
\def\jcp{J.~Chem.~Phys.}%
\def\jgr{J.~Geophys.~Res.}%
\def\jqsrt{J.~Quant.~Spec.~Radiat.~Transf.}%
\def\memsai{Mem.~Soc.~Astron.~Italiana}%
\def\nphysa{Nucl.~Phys.~A}%
\def\physrep{Phys.~Rep.}%
\def\physscr{Phys.~Scr}%
\def\planss{Planet.~Space~Sci.}%
\def\procspie{Proc.~SPIE}%
\bibliographystyle{mn.bst}
\bibliography{mybib.bib}
\end{document}